\RequirePackage{fix-cm}
\documentclass{svjour3}                     % onecolumn (standard format)
\smartqed  % flush right qed marks, e.g. at end of proof
\usepackage{xcolor}
\usepackage{hyperref}
\usepackage{booktabs}
\usepackage{breakcites}
\usepackage{subcaption}
\usepackage{paralist}
\usepackage{microtype}
\usepackage{multirow}
\usepackage{xspace}
\usepackage[flushleft]{threeparttable}
\usepackage{mathptmx}
\usepackage{mathtools}
\usepackage[scaled=0.92]{helvet}
\usepackage[scaled=0.85]{beramono}
\usepackage{enumitem}
\usepackage{appendix}
\usepackage{tikz}
\usetikzlibrary{arrows, shapes}
\usepackage{natbib}
\usepackage{blkarray}

\usepackage{graphicx}
\graphicspath{{./}}

\begin{document}

\title{Evolutionary Trends of Developer Coordination:\\A Network Approach}


\author{Mitchell Joblin \and
        Sven Apel \and
        Wolfgang Mauerer
}


\institute{Mitchell Joblin \at
              Siemens AG \\
              Wladmirstrasse 3, 91058 Erlangen, Germany \\
              Tel.: +49-176-61335168\\
              \email{mitchell.joblin.ext@siemens.com}           %  \\
           \and
           Sven Apel \at
           University of Passau \\
           Innstr. 33, 94032 Passau, Germany \\
           Tel.: +49-851-5093225\\
           \email{apel@uni-passau.de}
           \and
           Wolfgang Mauerer \at
           Technical University of Applied Science Regensburg \\
           Universit{\"a}tsstrasse 31, 93058 Regensburg, Germany \\
           Tel.: +49-941-9439753\\
           \email{wolfgang.mauerer@oth-regensburg.de}
}

\date{Received: date / Accepted: date}

\maketitle

\begin{abstract}
Software evolution is a fundamental process that
transcends the realm of technical artifacts and permeates the
entire organizational structure of a software project. By means of a longitudinal empirical study of 18 large open-source projects,
we examine and discuss the evolutionary principles that govern the coordination of developers.
By applying a network-analytic approach, we found that the implicit and self-organizing structure
of developer coordination is ubiquitously described
by non-random organizational principles that defy conventional
software-engineering wisdom. In particular, we found that: (a) developers form scale-free networks,
in which the majority of coordination requirements arise among an extremely small number of developers, (b)
developers tend to accumulate coordination requirements with more and more developers over time, presumably limited by an upper bound,
and (c) initially developers are hierarchically arranged, but over time, form a hybrid structure,
in which core developers are hierarchically arranged and peripheral developers are not. Our results suggest that
the organizational structure of large projects is constrained to evolve towards a state that
balances the costs and benefits of developer coordination, and the mechanisms used
to achieve this state depend on the project's scale.
\keywords{Software Evolution \and Developer Coordination \and Developer Networks}
\end{abstract}

\section{Introduction}\label{sec:intro}
Change in software is inevitable, and the constant pressure to adapt is a challenge that
all software projects encounter. 
%Adaptation pressures stem from assumptions embedded in the software's design and 
%implementation that often become invalid over time~\citep{lehman1997metrics}. 
%It has been shown that, if a project fails to respond
%to adaptation pressure, the degree of satisfaction provided by the software decreases with time~\citep{lehman1997metrics}.
The necessity of change is not isolated to the software design and implementation, it 
permeates through all artifacts and facets of a project including the entire \emph{organizational structure}. As the software evolves, the organizational structure building the software must 
also evolve to maintain effective coordination between developers. In the ideal case, a match or 
congruence is achieved between the coordination requirements implied by the project's 
technical artifact structure and the coordination mechanisms implied by the developer's 
organizational structure~\citep{Cataldo2008}.

The need for developer coordination is largely a consequence of software-artifact 
interdependencies. For example, two developers independently constructing coupled components 
must coordinate to avoid violating assumptions embedded in the components' design.
During software evolution, artifact interdependencies are added, 
removed, or changed, and local changes can propagate to dependent 
artifacts and alter the requirement for developers to coordinate. Without coherence between the
artifact structure and the organizational structure, developers
may lose awareness of new dependencies and their effects~\citep{Cataldo2009,Sosa2004}. Architectural documentation can help
support dependency awareness, however, it is difficult to maintain accurate documentation for an evolving system, and certain interdependencies may not be obviously expressed in the source code (e.g., code clones).  
Empirical evidence has demonstrated that the loss of interdependency awareness negatively influences
software quality and developer productivity~\citep{Espinosa2007,Cataldo2009,Cataldo2010}.

Our goal to understand the evolution of developer coordination
is motivated by specific scaling constraints known to affect software engineering.
One such scaling constraint arises from the quadratic relationship between 
team size and the number of possible interactions between developers. In
a group of $n$ developers, each developer can coordinate with $n-1$ other developers in
the group, such that the total number of possible coordination requirements is $n(n-1) = \mathcal{O}(n^{2})$. 
The implication is that, at a critical point, the overhead involved in coordination exceeds
the benefit of having additional developers~\citep{Brooks1975}. Consequently, the organizational structure of successful 
projects is constrained to evolve in a manner that mitigates the negative effects of
this and other
scaling constraints. We expect that the influence of the scaling constraints will
be observable through evolutionary trends of the organizational structure.

By gaining an understanding of the evolutionary patterns of developer coordination, software engineering practitioners
will be in a better position to identify and respond to changing coordination requirements.
It is well known that adding developers to a project often reduces overall productivity~\citep{Brooks1975,Scholtes2015},
but the precise mechanism behind this phenomenon is not yet well understood. It may be
the case that adding developers to a project negatively interferes with coordination requirements
by introducing additional complexity, which in turn causes
decreases in productivity. To address this unknown, we need to better understand the effect of adding developers on the
coordination structure. Once we have this understanding, we can establish processes for integrating new developers
that minimize the decrease of overall productivity by minimizing the influence to the existing coordination structure.

Methodologically, we use a network-analytic framework to conduct an exploratory study
on the evolution of developer coordination requirements. As developers complete their
tasks, their contributions to the source code may interact with the contributions
of other developers. To avoid unwanted side effects, developers must
coordinate their efforts to ensure their contributions interact
successfully according to the desired outcome.
To get a first-order approximation to the coordination requirements between
all developers in a project, we construct a \emph{developer network},
such that two developers are connected if they make contributions to
interdependent source code. We study the developer coordination structure, as embodied in a developer network,
with respect to the following three well-known and statistically
well-founded organizational principles:

\begin{compactitem}
\item \textbf{Scale freeness.} Scale-free networks are characterized by the existence of hub nodes 
with an extraordinarily large number of connections, which results in
several beneficial characteristics including robustness and scalability~\citep{dorogovtsev2013}. Developer networks of this kind are conjectured to
tolerate substantial breakdowns in coordination without significant consequences to software quality~\citep{dorogovtsev2013,Cataldo2010}.
\item \textbf{Modularity.} The local arrangement of nodes into groups that are internally well connected gives rise to a modular structure. Modularity is a notable characteristic of many complex systems and indicates the specialization of functional modules~\citep{dorogovtsev2013}. In the case of developer organization, this is the primary organizational principle used to reduce system-wide coordination overhead and increase productivity~\citep{Brooks1975}.
\item \textbf{Hierarchy.} The global arrangement of nodes into a layered structure, where small cohesive groups are embedded within larger and less cohesive groups, forms a hierarchy. Hierarchy is an organizational principle distinct from modularity and scale freeness, and has been shown to improve the coordination of distributed teams~\citep{Hinds2006}. For developer networks, hierarchy suggests the existence of stratification within the developer roles, and it indicates a centralized governance structure where decisions are primarily made at the top and passed down through a chain of command.
\end{compactitem}
\vskip 1em
\par
An important source of change in software projects is developer turnover, and this
phenomenon is likely to influence the evolution of the developer network's structural properties.
Open-source software projects are unique in that their organizational 
structure is predominantly self organizing, and they often lack a traditional
software-engineering process that supports coordination~\citep{dibona1999,Mockus2002,wolfgang}.
%Still, OSS proponents claim that the software 
%quality of OSS projects is equivalent
%or even superior, compared to commercially developed software~\citep{raymond1999}.
Conceptually, the lack of a centrally prescribed 
organizational structure enables open-source software projects to more easily adapt to 
evolutionary pressures~\citep{Sosa2004,kotter2014}.
One such evolutionary pressure is generated by \emph{developer turnover}: the process where
developers withdraw from a project and new developers join.
In open-source software projects, turnover exists in an extreme variety because the majority
of developers are peripheral and characteristically 
volatile~\citep{Mockus2002,Crowston05,koch2004}.
Typically, a high developer
turnover rate is portrayed as a severely detrimental circumstance
that poses a significant threat to the success of a software project primarily because of knowledge loss effects~\citep{Boehm89}. Curiously, in large open-source software
projects the harmonious coexistence of a large volatile peripheral group and a comparatively minuscule core group is ubiquitous~\citep{Mockus2002,Crowston2006,dtb05}. For this reason, we focus attention on
specific properties of the coordination structure that support the organization's
ability to benefit from the human resource of peripheral developers, which are
relatively abundant in comparison to core developers,
while at the same time, mitigating the risks implied by peripheral developers' volatility.
Specifically, we explore whether the structural features observed in the evolution of the developer network
enable open-source software projects to benefit from a large, but unstable, peripheral developer group. To clarify, we do not posit that turnover is a strictly positive or negative phenomenon, instead we direct our investigation to the relationship between developer turnover and the structural features of developer coordination.
Furthermore, we utilize the concept of core and peripheral developers to provide practical context for the explanation of evolutionary adaptation in the developer networks' structure. By gaining a better understanding of
how peripheral developers are embedded in the coordination structure relative to core developers,
we benefit from a number of practical insights. For example,
it may turn out that the more experienced developers (typically the core group)
and less experienced developers (typically the peripheral group)
are structurally arranged differently to help support effective coordination and transfer of knowledge between these distinct groups. 
%Still, it is astounding that such large groups of geographically
%distributed individuals can construct a complex, yet high-quality, software system 
%without explicit mechanisms to maintain a coordinated effort.

Capturing the evolution of developer coordination is challenging and demands
advanced techniques from software repository mining and time series analysis. In our
approach, we make use of sophisticated information retrieval methods to determine where
coordination requirements likely exist. We apply a sliding-window technique to transform
the discrete software changes recorded in the version-control system into a stream of
evolving developer networks. Finally, we analyze the stream of developer networks
with network analysis and statistical techniques to elicit insights into structural properties
in a time resolved manner.

By means of a longitudinal empirical study on the evolution of 18 popular open-source software projects,
we will address the following two main research questions~(RQ):
%
%and hypothesize that in the early phases 
%of a project a small set of developers form a centralized command and control structure.
%The result is a large amount of influence concentrated in a small 
%set of individuals, which also leads to a fragility in the event that a highly 
%influential person leaves. In the same vein, we expect that as a project matures, the 
%knowledge and 
%influence becomes distributed and likewise the command
%and control structure becomes more distributed and thus looses the original 
%hierarchical structure. Finally, modularity is recognized as a central 
%principle in reducing complexity in any system \TODO{add citation}.
\vskip 1ex
\noindent\textbf{RQ1: Change}---\emph{What evolutionary adaptations are observable in the history of long-lived open-source software projects concerning the three organizational principles?}
\vskip 1ex
\noindent\textbf{RQ2: Growth}---\emph{What is the relationship between properties of the three organizational principles and project scale?}
\vskip 1ex
As key results of our study, we found that developer networks are scale free when a project contains a large number of developers and primarily during temporal periods that coincide with project growth. With respect to modularity, developer networks become increasingly modular over time until an apparent upper bound is reached. Concerning hierarchy, developers form a hierarchical coordination structure in the early phases of a project, but over time the global hierarchy decomposes toward a state where only core developers are hierarchically arranged. Finally, we found that peripheral developers have significantly higher turnover rates compared to core developers, and the
adaptations observed in the developer coordination structure presumably help to mitigate
the risks implied by the abundant, but highly volatile, peripheral developer group.

\noindent
In summary, we make the following contributions:
\begin{compactitem}
\item We adapt a previously validated network-construction approach
to generate and analyze a temporally-ordered sequence of evolutionary changes that occur in developer networks by applying a sliding-window technique to extract historical
data from version-control systems.
\item We address the research questions of our study by means of detailed analyses of the entire publicly available history of 18 popular open-source software projects
with long and complex histories, some of which date back more than 23 years.
The infrastructure we developed to perform the analysis is publicly available.
\item We demonstrate that developers form organizational structures that are statistically improbable to occur in purely random networks,
which indicates the presence of non-random organizational principles.
\item We identify and discuss a number of general evolutionary trends that describe how developer networks
evolve in software projects with respect to scale freeness, modularity, and hierarchy.
\item We present empirical evidence that groups of core developers are significantly more stable than groups of peripheral developers and we discuss how the developer network's structural features accommodate
these two distinct developer groups.
\end{compactitem}
\vskip 1ex
\noindent
In addition to these contributions, we improve on a number of known methodological deficiencies in prior empirical studies noted by~\citet{crowston2012}. Specifically, the deficiencies include: drawing conclusions from the analysis of very few (often only one) projects, a lack of attention to the early transitional phases by focusing primarily on successful projects after they have become well established, the use of coarse-grained data to draw conclusions, and a general lack of longitudinal studies that explore the different phases of evolution.

All experimental data and source code are available at a supplementary Web site:\\
\url{http://siemens.github.io/codeface/emse/}\,.

\section{Background}\label{sec:ev_background}
In the following section, we discuss approaches for abstracting the technical activities of developers as a network that reflects coordination requirements among developers. We then introduce the conceptualization of core and peripheral developer roles and conclude
with an introduction to the concepts
and definitions of scale-free networks, network modularity,
and network hierarchy.

\subsection{Developer Networks}\label{sec:developer_networks}
In social sciences, networks are frequently used as a mathematically convenient structure to study the relationships
among a set of actors involved in a mutual activity that is social in nature. In this sense, developer networks are socio-technical networks,
where the mutual activities are technical in nature, stemming from the software-development process, but still imply relationships between people.
The purpose of constructing a developer network is to obtain an authentic representation of
the developer-coordination requirements implied by their development activities~\citep{Meneely2011,cataldo2006identification,joblin2015,Begel2010}.
A common approach is to construct a developer network based on the developers'
contributions to software artifacts by extracting operational data from a version-control system~\citep{joblin2015,Lopez-fernandez2006,martinez2008using,Jermakovics2011}.
In this type of network, the nodes represent developers and edges placed
between developers represent mutual contributions to common technical artifacts.
One drawback of these approaches is that they assume all lines of code within a single artifact are uniformly interrelated and any relationship between related code that is scattered across multiple artifacts is neglected. Below, we discuss several possible heuristics---with
varying degrees of precision---to determine when two developers likely have a coordination requirement. We emphasize that all known approaches rely fundamentally on heuristics and therefore only provide indications of where coordination requirements exist. Occasionally, developers are incorrectly portrayed as having a coordination requirement when, in fact, they do not, or legitimate coordination requirements fail to be represented. For this reason, it is useful to adopt a probabilistic perspective on developer networks by interpreting the network edges in terms of a likelihood that developers exhibit a coordination need. An example of such a perspective is, on average, pairs of developers connected by an edge are more likely to exhibit a coordination requirement compared to pairs of developers not connected by an edge.

\begin{figure}[t]
\centering
\includegraphics[width=0.95\linewidth]{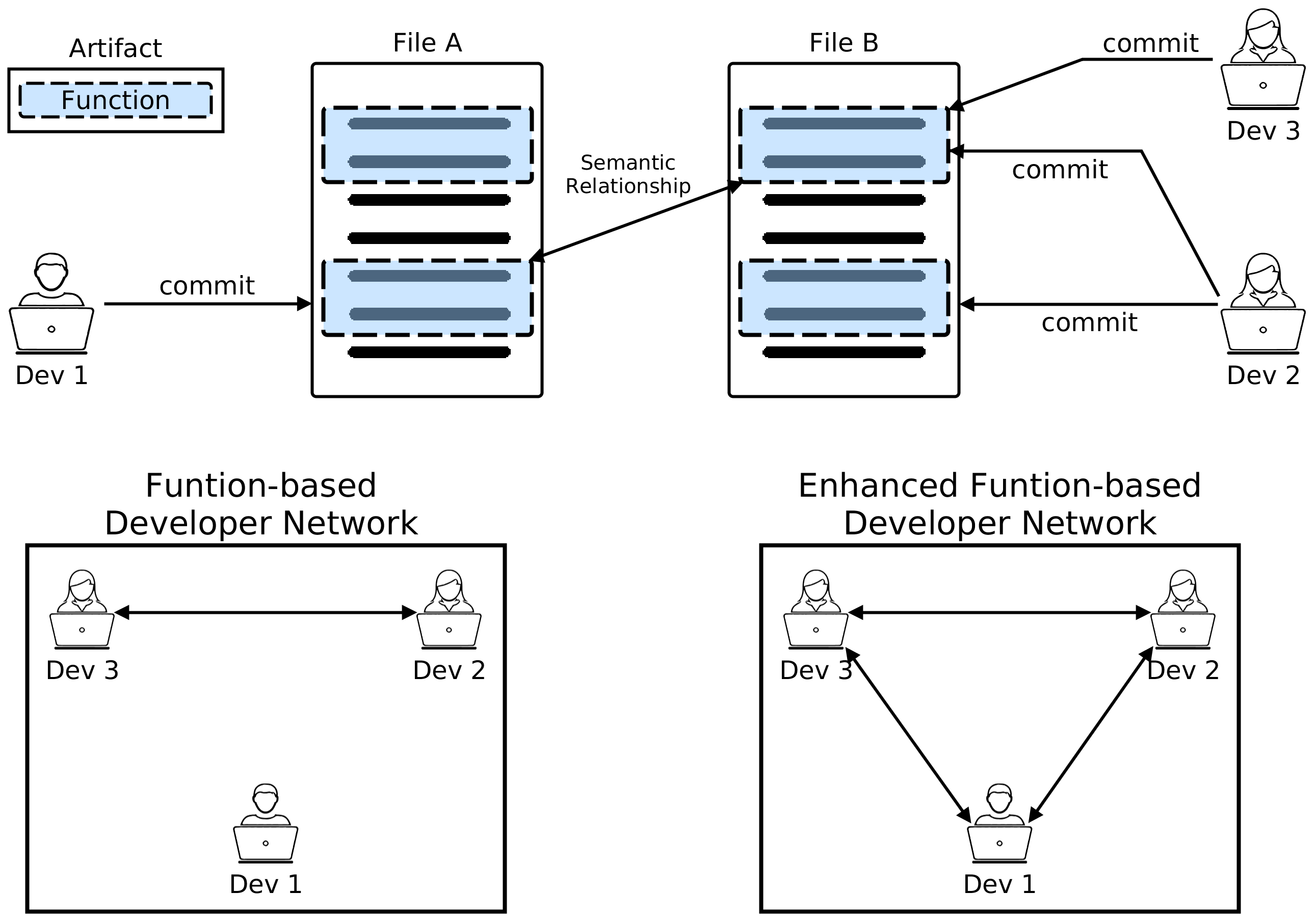}
\caption{Three developers edit two semantically coupled functions in separate files (top). The resulting developer network from applying the original function-based construction method is shown bottom left.
The resulting developer network from applying our enhanced construction approach that includes the coupling between artifacts is shown bottom right.}
\label{fig:dev_network}
\end{figure}

\subsubsection{Artifact Contribution}\label{sec:artifact_contribution}
The most popular family
of heuristics is based on contributions to a common artifact (e.g., files or functions)~\citep{Lopez-fernandez2006,Jermakovics2011,martinez2008using,Meneely2008,Meneely2011,joblin2015}.
The rationale is that an artifact is an abstraction of the software system that represents
a bundle of cohesive functionalities, and developers are constrained to coordinate by virtue of the interdependencies
that exist among the lines of code that compose the artifact. The chosen granularity of the artifact has implications on the authenticity of the resulting
developer network. For example, a coarse granularity (e.g., files) results in identifying more relationships
between developers, but can decrease accuracy by identifying relationships that are not reflective of reality.
In contrast, a fine granularity (e.g., functions) will increase accuracy,
but may omit some developer relations. In the bottom left portion of Figure~\ref{fig:dev_network},
we illustrate an example of the fine-grained heuristic for identifying developer coordination requirements using contributions to common functions.
This particular heuristic has shown promise in constructing accurate developer networks and was validated by surveying open-source developers regarding the network's correctness with respect to reflecting the developers' perception~\citep{joblin2015}.

It is important to recognize that, in all these approaches, an artifact is simply a means of grouping lines of code that are assumed to be interrelated. It can be the case, especially for large artifacts (e.g., God classes)~\citep{ocb09}, that many of the lines are in fact not closely related. Additionally, lines of code can also exhibit meaningful relationships that transcend the artifact boundaries. An option to overcome these challenges is to choose a fine-grained artifact (e.g., functions), and then have a means to identify meaningful artifact relationships that cross the artifact boundary (e.g., using a notion of artifact coupling).  
%We further enhanced the
%original method using artifact coupling information to construct a more %complete network because the original approach misses some coordination %relationships. An example of the differences between our enhanced %approach and the original approach for computing function-based %developer networks is shown in Figure~\ref{fig:dev_network}. 

\subsubsection{Artifact Coupling}\label{sec:artifact_coupling}
Software systems are intrinsically coupled with complex associations between their artifacts~\citep{arias2011practice}.
Through artifact coupling, developer tasks become interdependent, because changes to one artifact
can propagate along the coupling relationship, and developer coordination is required to manage unintended ripple effects~\citep{arnold1993impact}.
For this reason, it is beneficial to incorporate the common understanding of software-impact analysis~\citep{arnold1993impact} to support the identification of developer-coordination requirements by augmenting
developer networks with software-coupling information.
The upper portion of Figure~\ref{fig:dev_network} illustrates the common situation where developers
contribute to code that is related, but contained in separate functions in separate files.
As shown in the bottom left of Figure~\ref{fig:dev_network}, without considering the relationship
between the artifacts (as in the function-based heuristic of Section~\ref{sec:artifact_contribution}) , the coordination requirement is missing between developer 1 and the remaining developers.
By augmenting the developer network with coupling information, one is able to recover the missing coordination requirement,
as shown in the bottom right portion of Figure~\ref{fig:dev_network}.

A diverse set of coupling mechanisms exist in software systems
(e.g., function calls, class inheritance, data dependencies etc.), but
prior research has shown that not all coupling mechanisms are equally reflective of developer perception~\citep{bavota2013}.
Coupling relationships that reflect the developer's mental model are important
for coordination purposes, because developers will inherently rely
on their internal understanding of the system to coordinate their
work with others. Essentially, if a developer perceives their changes to influence another developer's work, or vice versa, it is more likely for them to recognize a need to coordinate their efforts than if there is no expectation of impact. Empirical research indicates that
traditional static and dynamic coupling mechanisms do not closely resemble developer perception~\citep{bavota2013},
and, the implications of many traditional coupling mechanisms are understood by the developer because they are explicitly expressed in the source code~\citep{Cataldo2009}.
When the implications of a dependency are known then it is less critical
that developers coordinate because the source
code serves as a means to avoid incorrect assumptions about the code.
\citet{bavota2013} have shown that semantic coupling is more indicative of coordination requirements because artifact relationships at the semantic level
are more likely to reflect the developer's mental model and the implications
are less likely to be fully understood by other developers than structural or dynamic
dependencies.

Techniques for extracting coupling relationships that make use of information-retrieval methods have shown promise in raising the coupling concept to a semantic level that agrees with developer perception~\citep{bavota2013}.
From a purely practical perspective, semantic coupling is also well suited,
because it is programming-language independent, allowing the comparison between projects implemented in different programming languages.
Semantic coupling is based on the principle that domain knowledge is embedded
in the textual content of the software's implementation artifacts (i.e., variable identifiers, function names, parameter names, comments, etc.), and artifacts implementing related domain concepts will
share a common vocabulary~\citep{poshyvanyk2009using}. For the purpose of constructing developer networks, semantic coupling serves as an alternative means to group code together on the basis of the domain concept they concern. At a high level, the challenge of quantifying semantic coupling depends on the identification of key terms from the overall implementation vocabulary that can be used to discriminate between distinct domain concepts. \emph{Latent semantic indexing} is often applied together with a term-weighting step called
term-frequency inverse document frequency (TF-IDF) to extract the semantic
coupling information~\citep{Manning2008}. Similarity between documents is measured using cosine similarity in the latent space, to determine whether two artifacts are semantically coupled according to a certain threshold~\citep{poshyvanyk2009using}. A More detailed discussion of these topic is provided in Appendix A.

\subsection{Core and Peripheral Developers} \label{sec:core_peri}
All software projects face the situation that developers withdraw at some point and need to be replaced by new, often less experienced, developers. This process of \emph{developer turnover} can present enormous risks to software projects, because crucial knowledge is often lost with departing developers~\citep{Boehm89,Mockus2010,huselid95}. Another consequence of developer turnover is that new developers initially require mentorship, thereby consuming additional human resources by placing a burden on more experienced developers in the project. This is one factor that contributes to the well-known phenomenon that adding developers to a late project causes further delays~\citep{Brooks1975}. In open-source software projects, developer turnover exists in an extreme variation because the vast majority of developers have occasional, short-term participation, and generally only a very small number of core developers have consistent long-term participation~\citep{Mockus2002,Crowston05,koch2004}.
It is extraordinary that open-source software projects are able to thrive under the extreme conditions of high developer turnover. For this reason, we dedicate attention to study the evolutionary pressures caused by the significantly different turnover rates between core and peripheral developers.

\emph{Core developers} are characterized by prolonged, consistent, and intensive participation in the project, and they often have extensive knowledge of the system architecture and strong influence on project decisions~\citep{Mockus2002}. In contrast, \emph{peripheral developers} are characterized by irregular, and often short-lived, participation in the project. The peripheral developer group is the larger of the two, by a significant margin, but core developers are responsible for doing most of the work~\citep{Mockus2002,Crowston05}. While peripheral developers are an abundant human resource, they also introduce risk and consume resources. For example, empirical evidence indicates that changes made by peripheral developers introduce more architectural complexity than changes made by core developers~\citep{Terceiro10}. Researchers have shown that developer turnover negatively impacts code quality, in terms of bug density~\citep{foucault2015}. Therefore, a stable and knowledgeable core developer group is imperative for ensuring system integrity in the presence of potentially inadequate changes introduced by peripheral developers. However, it appears to be ubiquitously true that successful open-source software projects are capable of benefiting from a large number of volatile peripheral developers, while at the same time, mitigating the associated risks. Since the coordination structure of popular open-source software projects supports the symbiotic coexistence of a highly volatile peripheral developer group and a comparatively stable core developer group, we expect to observe adaptations in the developer network that promote system integrity and effective coordination.

Metrics used to classify a developer as core or peripheral generally quantify the amount of participation a developer has in the project, such as lines of code or number of commits contributed~\citep{Terceiro10,Crowston05,robles09}. A developer is then assigned to the core group if their level of participation is in the upper 20th percentile; all other developers are considered to be peripheral~\citep{Terceiro10,robles09}. Metrics based on lines of code and commits pose significant threats to validity. Since trivial whites space changes, moving code from one file to another, or code-style changes are all counted the same as new feature code, certain developers may artificially appear to be extremely active. With respect to using commits for classifying developers, empirical studies have shown that the variance in commit size is extremely large in open-source projects and so all commits should not be considered equal~\citep{arafat2009commit}. An alternative approach is to operationalize developer roles using social-network analysis concepts, such as degree centrality. Core developers are individuals that are highly central (e.g., high degree nodes) in the developer network and peripheral developers are individuals that are not highly central (e.g., low degree nodes). A recent study involving 166 open-source developers found that, while many of the core--peripheral metrics are to some extent consistent with each other, developer network-based metrics (e.g., degree and eigenvalue centrality) reflect developers' perception of roles most accurately~\citep{joblin2016}.

\subsection{Scale-Free Networks}\label{sec:scale_free_networks}
\begin{figure}[!t]
	\centering
	\begin{tabular}{ccc}
		\textbf{Exponential} & \hspace{0.25cm}
		\textbf{Power-Law} \\
		\includegraphics[width=0.4\textwidth]{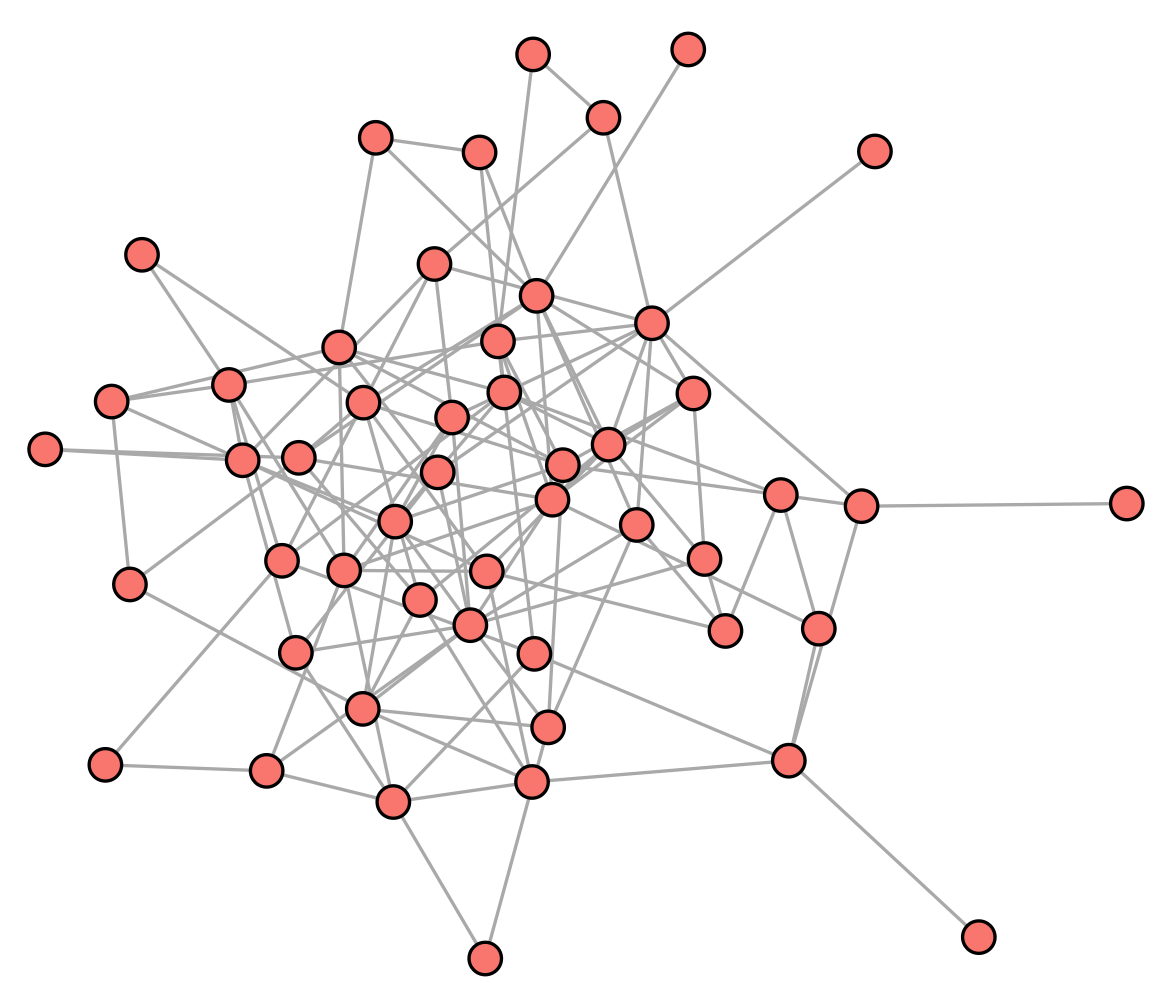} & \hspace{0.5cm}
		\includegraphics[width=0.4\textwidth]{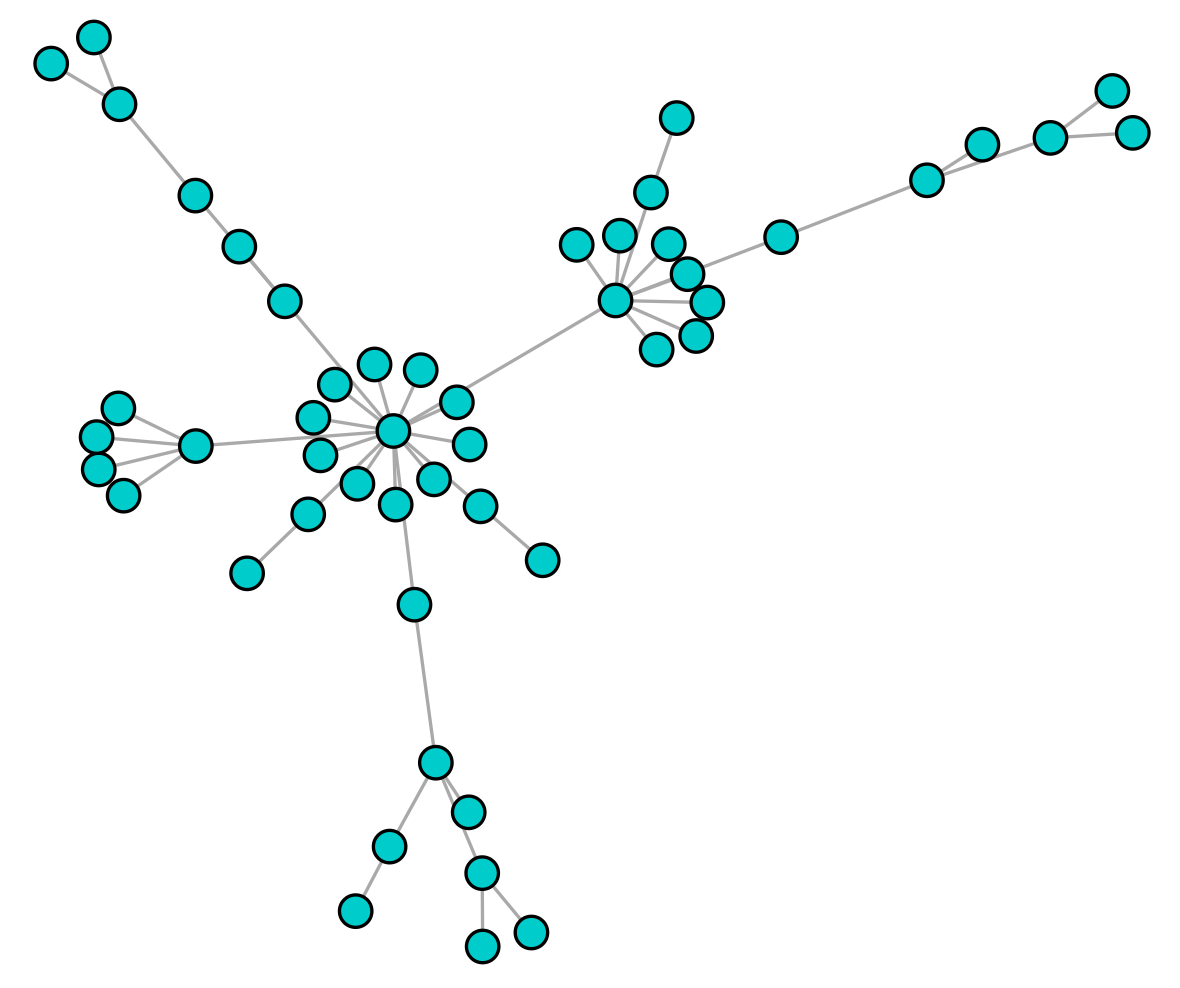} \\
		&  & \\
		\multicolumn{3}{c}{\includegraphics[width=0.85\linewidth]{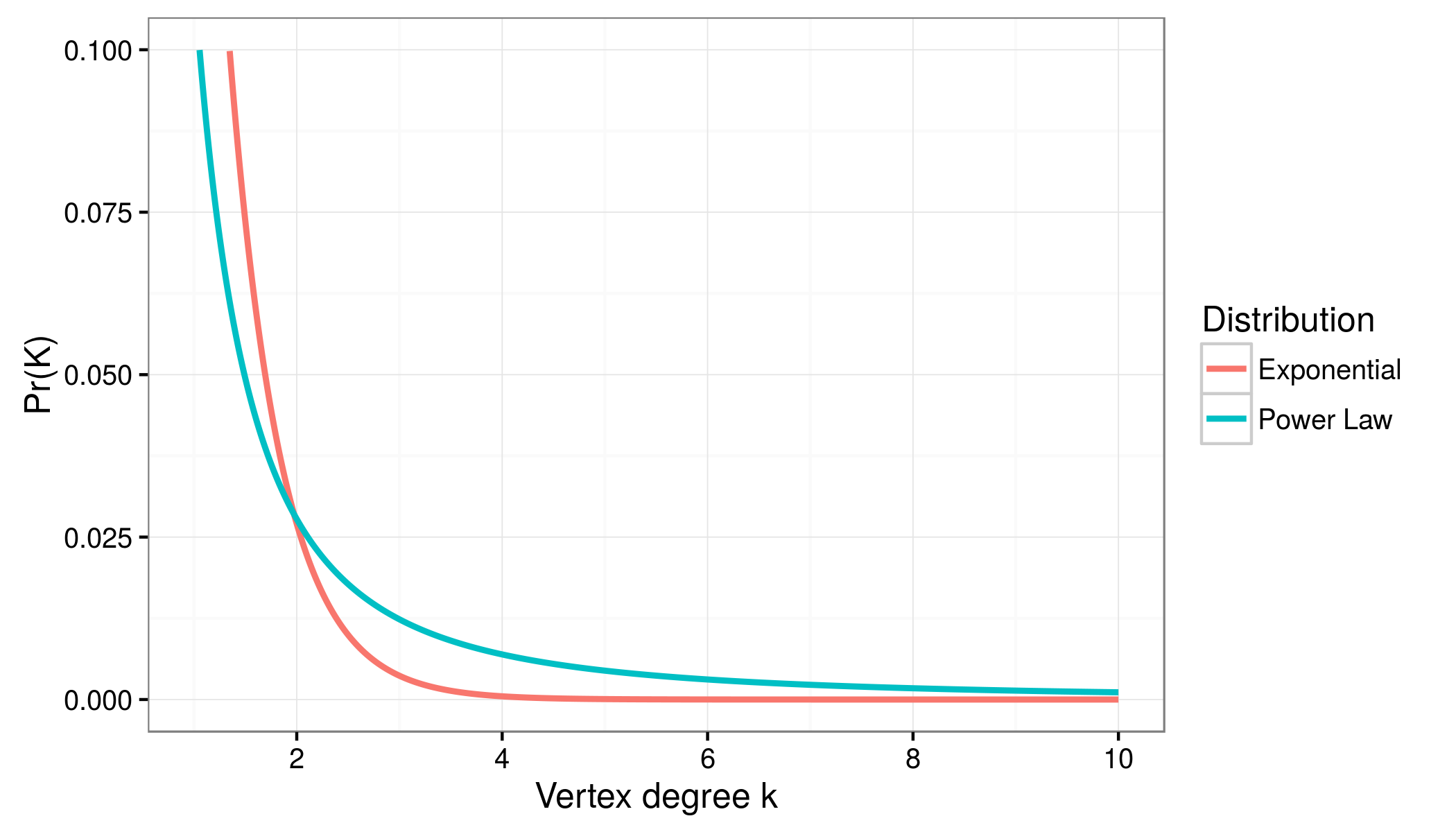}}
	\end{tabular}
	\caption{An ER random network with an exponential degree distribution (top left) and scale-free network with a power-law degree distribution (top right). The corresponding degree distribution for exponential $\mathrm{Pr}(k;\alpha)=e^{-\alpha k}$ and power-law $\mathrm{Pr}(k;\alpha)= k^{-\alpha}$ are shown (bottom). The power-law distribution contains significantly more weight in the right tail compared to the exponential distribution. The heavy tail gives rise to the organized structure of the network where a small number of nodes are hubs and low degree nodes collect around these hubs.}
	\label{fig:degree_dist_expl}
\end{figure}

Early research characterized the topology of complex networks according to the Erd\H{o}s-R\'{e}nyi (ER) model for random graphs, in which
edges are independent and identically distributed with a fixed probability~\citep{erdos1959}. 
In the ER model, the network edges are distributed according to a binomial distribution and the degree distribution is Poisson (i.e., exponential). This results in
the existence of a ``typical'' node that is representative of most nodes in the network, 
and it is extremely rare to observe ``hub'' nodes with significantly more connections.
More recently it has been shown that many real-world networks from fundamentally different
domains (e.g., biology, sociology, scientific paper authorship, Internet 
routers) exhibit a substantially more organized structure than initially expected~\citep{jeong2000,bernard1988,barabasi2002,dorogovtsev2013}.
This class of networks obeys a power-law degree distribution and are referred to as ``scale-free''.
In this model, there is no notion of a typical node; hubs with many more 
connections than the average are common. \emph{Scale-free} networks exhibit a
structure that is the product of organizing principles that are far from
uniform randomness. In the literature, it is hypothesized
that scale-free networks grow according to the organizational principle of \emph{preferential attachment},
according to which nodes entering the network have a bias to attach to already well-connected nodes~\citep{barabasi1999}. Although, preferential attachment is only one of many possible explanations for the formation of scale-free networks~\citep{dorogovtsev2013}, it could explain the evolution of open-source software projects, because it is plausible that new developers with little experience require mentorship from highly knowledgeable core developers or core developers supervise important parts of the system, which make
it necessary for peripheral developers to coordinate with them. These conditions would then
lead to a situation of preferential attachment and result in a scale-free network.

We are particularly interested in the scale-freeness property of developer networks because it has a number of beneficial
characteristics, including robustness to perturbations~\citep{dorogovtsev2013}. This means that a random removal of a node is unlikely to disturb the connectivity of the network (e.g., fracturing the network into isolated subgraphs). In the case of a software project, robustness indicates that the withdrawal of a random developer should not severely destroy the topology of the network (i.e., the organizational structure). However, it is important to recognize that a network can only be optimized to be robust for a particular removal mechanism. As mentioned, scale-free networks are extremely robust to \emph{random removals}, but the compromise is that they are extremely vulnerable to targeted removals. That is to say, a removal of only a small number of hub nodes can completely destroy the network topology. The question is then, does a scale-free topology offer beneficial characteristics for an open-source software project? The answer to this question depends on the relative likelihood for withdrawal of core developers (i.e., hub nodes) and peripheral developers (i.e., low degree nodes). If core developers are indeed less likely to leave the project compared to peripheral developers, then a scale-free topology is beneficial since the removal process does not target hub nodes. We will specifically examine these relative turnover rates in core and peripheral developers to determine whether a scale-free network increases the robustness of the project's structure.

To identify a scale-free network, one must show that the degree
distribution is plausibly described by $\mathrm{Pr}(k;\alpha) \propto k^{-\alpha},$
where $\mathrm{Pr}(k;\alpha)$ is the probability of observing a node with $k$ connections and
$\alpha$ as the power-law scaling parameter. In Figure~\ref{fig:degree_dist_expl}, we illustrate the differences between an ER random graph and a scale-free network and show the influence that a power-law degree distribution has on the network topology. In Section~\ref{sec:scale-free}, we will discuss details of the technique for determining whether an observed network is scale free.

\subsection{Network Modularity}\label{sec:network_modularity}
\begin{figure}[t]
	\centering
	\includegraphics[width=0.9\linewidth]{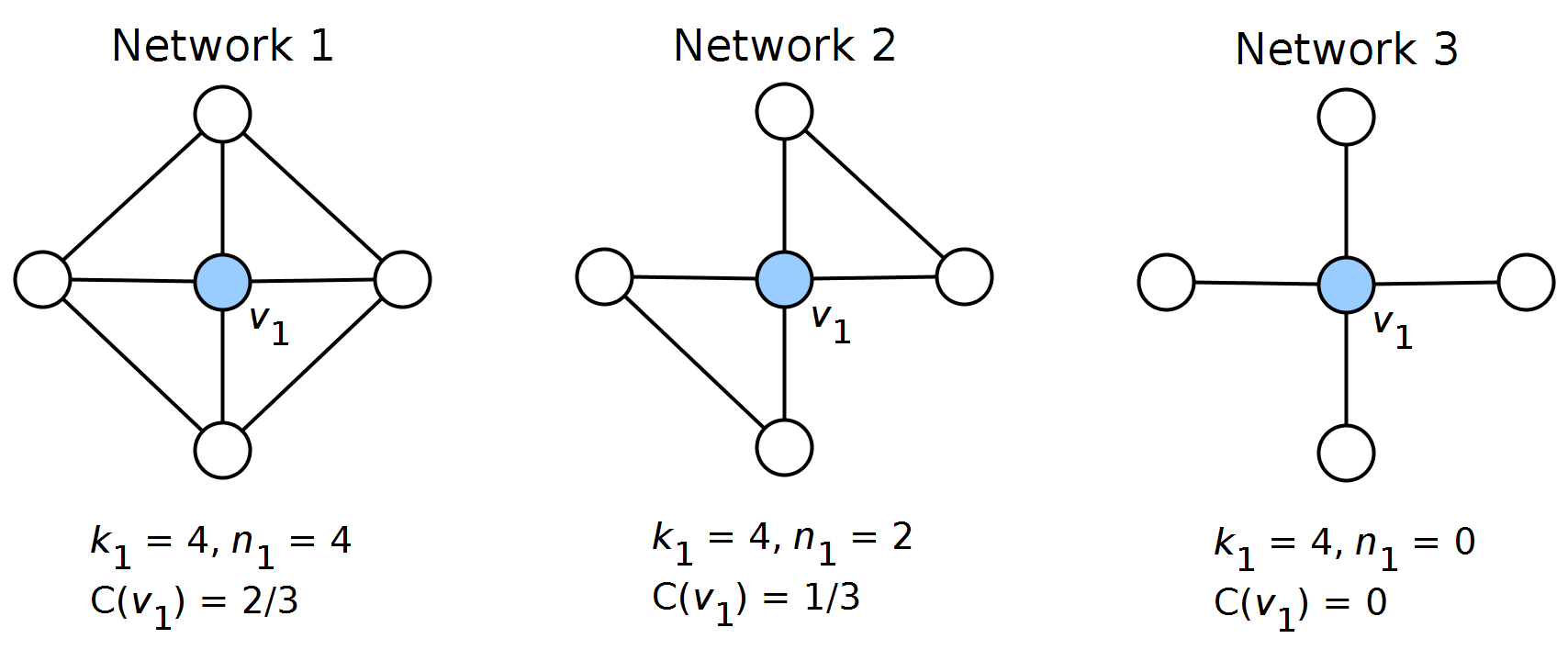}
	\caption{Three networks illustrating decreasing clustering coefficients (left to right) for the node labeled $v_1$.
		In network 1, the clustering coefficient is highest, because many neighbors of $v_1$ are connected. The neighbors
		of $v_1$ in network 2 are less connected compared to network 1, thus reducing the clustering
		coefficient of $v_1$. In network 3, none of the neighbors are connected and therefore the clustering coefficient is zero.}
	\label{fig:cc_expl}
\end{figure}
The scale-freeness property is a statement about individual nodes and their respective
degrees, but it entirely neglects features regarding the connectivity of a node's local neighborhood. In our case, the local neighborhood
of developer $X$ is a subnetwork that represents all coordination requirements between all developers that are connected to developer $X$. 
\emph{Modularity} captures the connectivity of a node's local neighborhood by quantifying the extent to which nodes
form connected groups. In the analysis of software-artifact relationships,
coupling (the extent to which an artifact is externally dependent) and
cohesion (the extend to which an artifact is internally dependent) are
frequently used.
Modularity is expressed as a relationship between the concepts of
coupling and cohesion such that a
highly modular structure is one that exhibits low coupling and high cohesion~\citep{smc74}.
In a social network, modularity is high when the neighbors of node $i$ have relations 
to other neighbors of node $i$, which is called a \emph{triadic closure}. 
The conjecture is that, for three nodes $X$, $Y$ and $Z$,
edges $(X, Y)$ and $(X, Z)$ increase the likelihood for edge $(Y, Z)$ to exist by virtue of the
commonality from both $Y$ and $Z$ being connected to $X$. This natural clustering has been shown to exist
in many real-world networks (e.g., it indicates specialization of function in biological networks or
people with common interests in social networks)~\citep{dorogovtsev2013}. 
Similarly, in developer networks, we expect modularity to arise from
specialization in the developer roles and contributions to interdependent tasks. 
This expectation follows from Conway's law, which
hypothesizes that the modular structure of a software system should be
reflected in the developer organization~\citep{Conway1967}. 

To quantify modularity, we use the well studied \emph{clustering coefficient}:
\begin{equation}
c_{i} = \frac{2 n_{i}}{k_{i}(k_{i} - 1)},
\end{equation}
where $n_{i}$ is the number of edges between the $k_{i}$ neighbors of node $i$~\citep{boccaletti2006complex}.
The intuition is that $k_{i}(k_{i} - 1) / 2$
edges can exist between $k_{i}$ nodes, and the clustering coefficient is a ratio that reflects the fraction of existing
edges between neighbors divided by the total number of possible edges. For example, if a node has a high clustering coefficient,
then many edges exist between the neighbors of this node. Conversely, if a node has a low clustering coefficient, then
only a few edges exist between neighbors of this node. In Figure~\ref{fig:cc_expl},
three networks are shown in which the solid node has a decreasing clustering coefficient
from left to right.

\subsection{Network Hierarchy}\label{sec:network_hierarchy}
\begin{figure}[!t]
	\centering
	\begin{tabular}{ccc}
		\textbf{Random Network} & \hspace{0.5cm} &
		\textbf{Hierarchical Network} \\
		\includegraphics[width=0.35\textwidth]{er_network} & \hspace{0.5cm} &
		\includegraphics[width=0.35\textwidth]{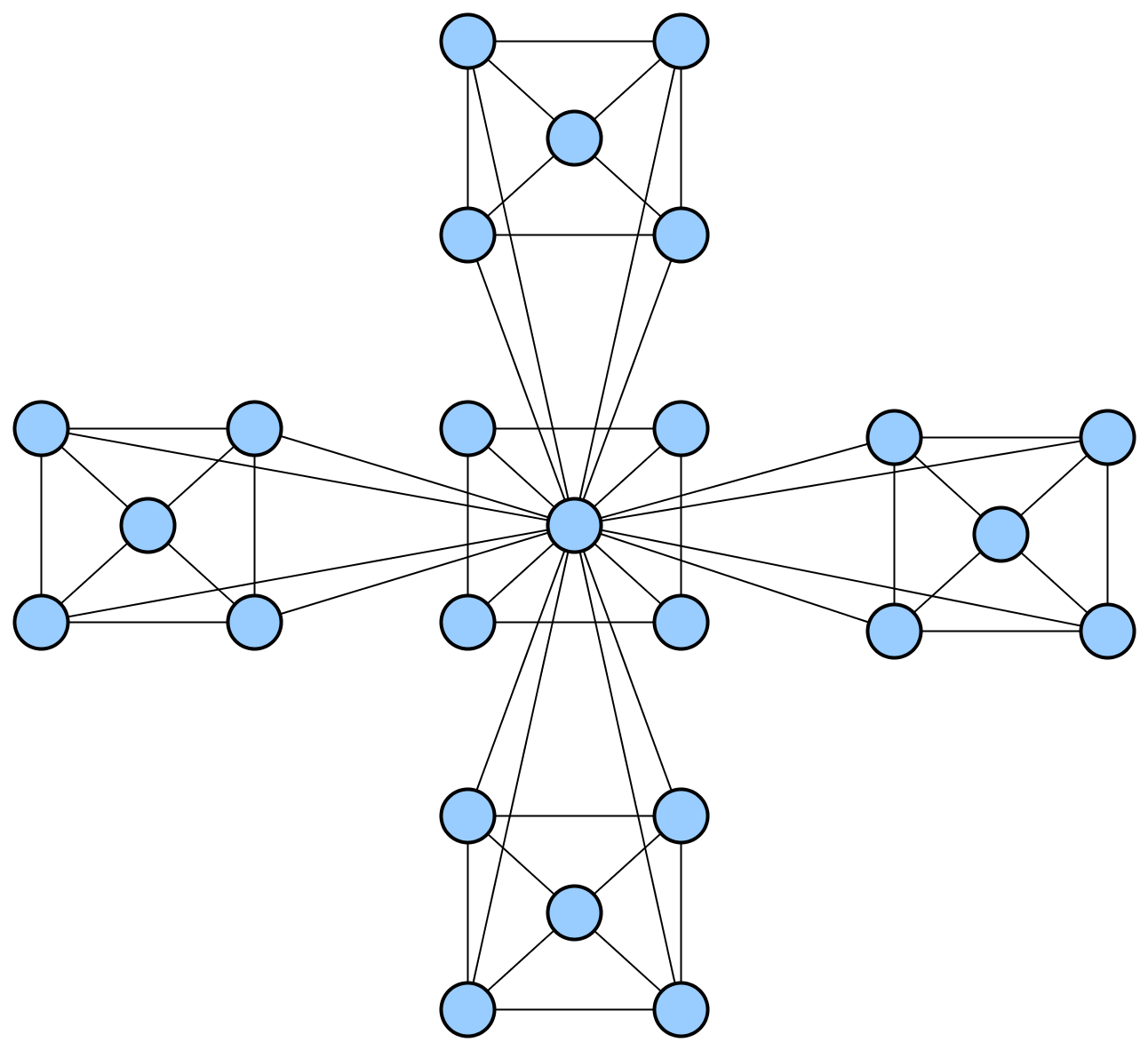} \\
		& & \\
		\includegraphics[width=0.4\textwidth]{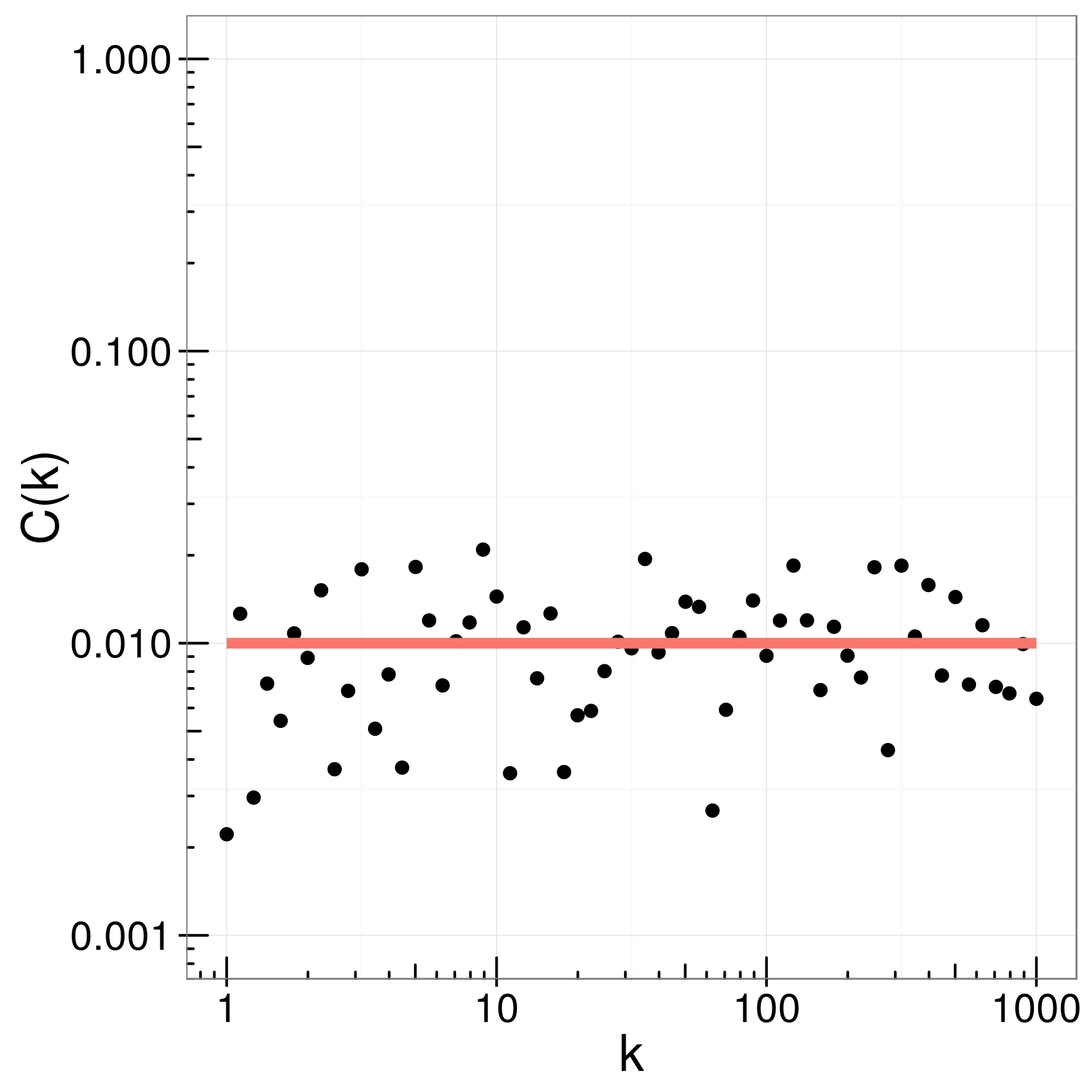} & \hspace{0.5cm} &
		\includegraphics[width=0.4\textwidth]{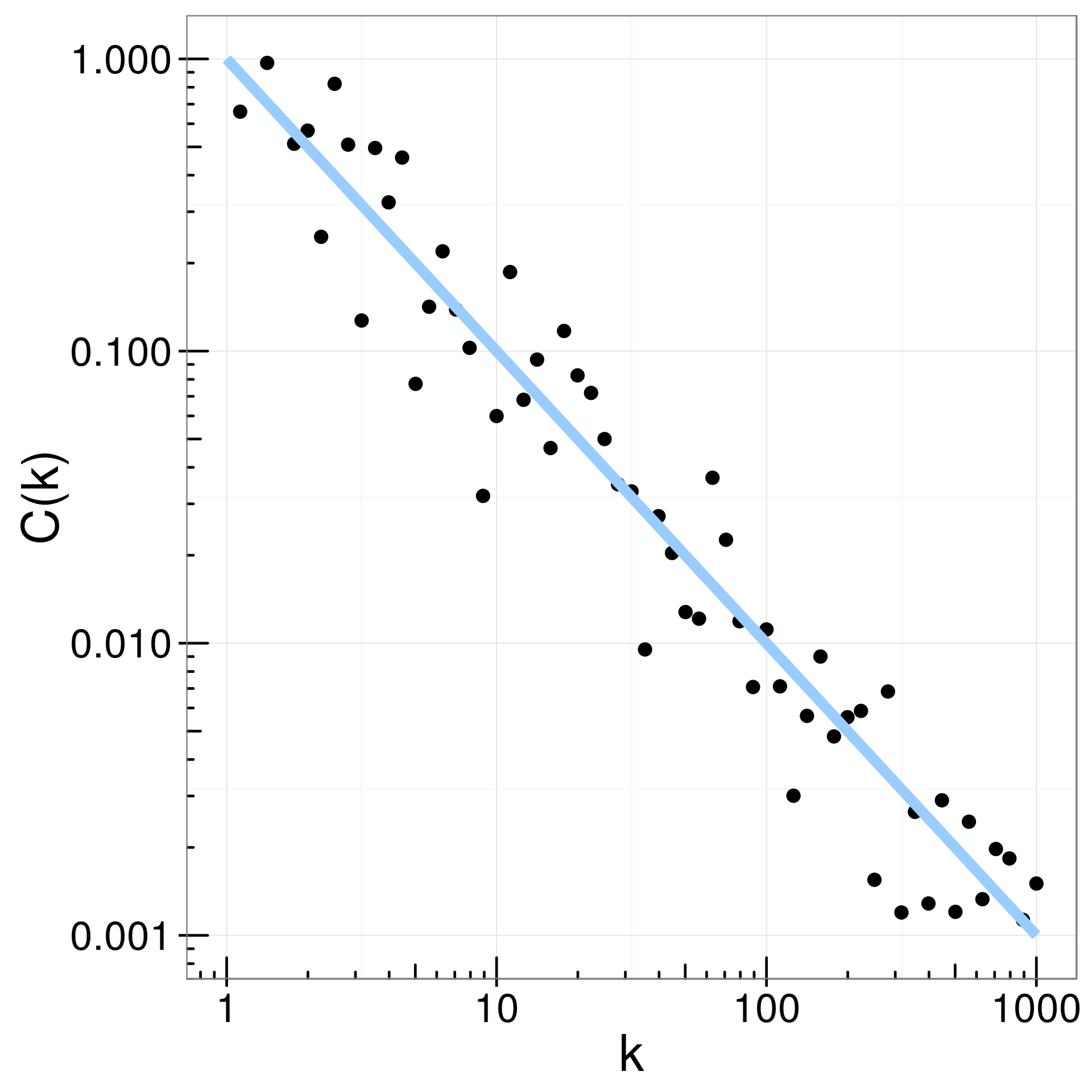}
	\end{tabular}
	\caption{ER random network (left) and hierarchical network (right) with corresponding scatter plot of node degree $k$ versus clustering coefficient $C(k)$. The hierarchical network topology deviates from randomness by having small cohesive clusters that are embedded withing larger and less cohesive clusters. The hierarchical topology manifests as a linear dependence between node degree and clustering coefficient, illustrated by a line with negative slope, which is not present in ER random networks.}
	\label{fig:hierarchy_expl}
\end{figure}

So far, we have introduced scale freeness, which describes the distribution of edges
among nodes, and network modularity, which describes the grouping of nodes according to the local network
structure. The concept of \emph{hierarchy} brings these two concepts together by addressing how 
local groups are arranged relative to each other. In a hierarchical network, there exists stratification within the
network that stems from cohesive groups being embedded within larger and less cohesive groups. This stratification
is manifested as a relationship between the node clustering coefficient and the number of connections, that is, the node degree~\citep{ravasz2003hierarchical}. 
Nodes with high degree and low clustering coefficient
represent the top of the hierarchy; nodes with low degree
and high clustering coefficient are located at the bottom of the hierarchy. 
The relationship between node degree and clustering coefficient in a hierarchical network is described by
$C(k) \propto k^{-1},$  
where $C(k)$ is the clustering coefficient for a node with degree $k$~\citep{ravasz2003hierarchical}. In Figure~\ref{fig:hierarchy_expl}, the difference between the topology of a hierarchical network and ER random network are shown together with the corresponding relationship between clustering coefficient and node degree.

To test for the presence of hierarchy, a statistically sound approach is to apply a robust linear regression technique to solve for the optimal linear model satisfying the functional form $Y = \beta_{0} + \beta_{1}X$, where the clustering coefficient is the response variable denoted by $Y$ and node degree is the predictor variable denoted by $X$. If the optimal
linear model has a nonzero slope (i.e., $\beta_{1} < 0$) and the slope parameter is statistically different from zero, such that $p < 0.05$, where $p$ is that probability that $\beta_{1} = 0$, we can conclude that hierarchy is present.

Intuitively, the hierarchical relationship between node clustering coefficient and node degree implies that nodes of high degree tend to be connected to many different
groups that are themselves loosely coupled to each other, meanwhile low degree nodes form highly connected groups. What makes hierarchy particularly interesting
is that it is not explained solely by preferential attachment and therefore indicates an entirely
separate organizational principle~\citep{ravasz2003hierarchical}. 
Hierarchy in developer networks indicates the existence of an organizational structure that
transcends the local network structure and represents an organizational element that sprawls the network topology at different
layers of abstraction to improve coordination between developers that are members of different groups~\citep{Hinds2006}.

\section{Methodology}
We now discuss our empirical methodology to study
the evolutionary principles of developer coordination
in open-source software projects. The study is divided into two parts:
First, we construct a series of developer networks
using historical data stored in the version-control systems of a set of
subject projects. Second,
we apply network-analysis techniques to examine the network topology
(scale-freeness, modularity, and hierarchy) as a function of time and relate
the measurements to developer turnover rates in the groups of core and peripheral developers.

\subsection{Developer Network Construction} \label{sec:network_construction}

\begin{figure}[t]
	\centering
	\includegraphics[width=0.5\linewidth]{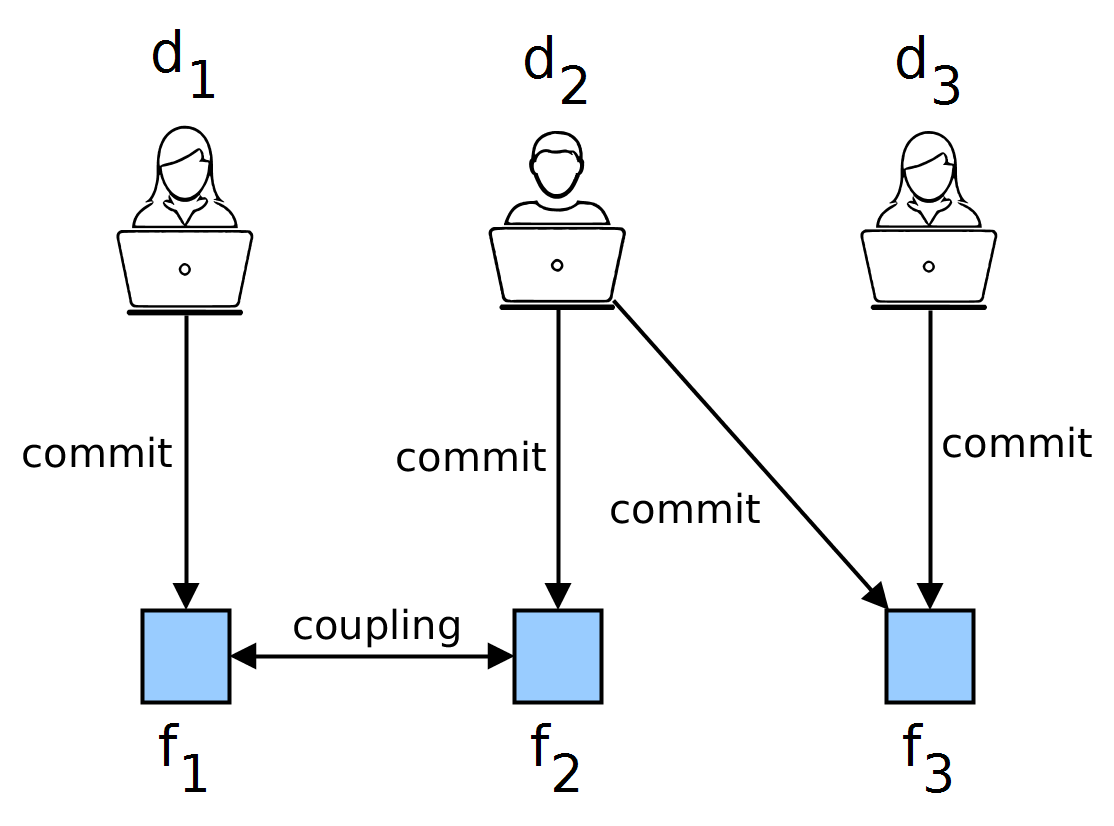}
	\caption{Three developers make commits to three different functions.
		Functions $f_{1}$ and $f_{2}$ are shown to have a coupling relationship between them.}
	\label{fig:network_expl}
\end{figure}

Our approach is inspired by the framework proposed by~\citet{Cataldo2008}
for identifying coordination requirements between developers working on interrelated tasks. To capture a more complete model of the developer coordination (see Section~\ref{sec:artifact_contribution}), we improve on the state of the
art~\citep{Lopez-fernandez2006,Jermakovics2011,martinez2008using,Meneely2008,Meneely2011,joblin2015} by
augmenting developer networks with semantic artifact-coupling information.
While our implementation is one variation that uses a function-level artifact and semantic coupling,
this approach naturally extends to accommodate other artifacts
(e.g., configuration files, requirements, documentation etc.) and coupling mechanisms
(e.g., dynamic, structural, co-change etc.). Figure~\ref{fig:dev_network} provides an
illustration of the network-construction approach. Specifically, we reconcile the developer-artifact
contribution (see Section~\ref{sec:artifact_contribution}) and software-artifact coupling (see Section~\ref{sec:artifact_coupling})
information as follows: We begin by first identifying all developers' contributions to all functions in the system and express the contributions of $M$ developers to $N$ functions in an $M\times N$ matrix as
\begin{align}
A_{\text{contrib}} &= 
\begin{bmatrix}
f(d_{1}, a_{1})&  \hdots & f(d_{1}, a_{N})\\ 
 \vdots & \ddots & \vdots \\ 
f(d_{M}, a_{1}) & \hdots & f(d_{M}, a_{N})
\end{bmatrix},
\end{align}
where $A_{\text{contrib}}$ is the function-contribution matrix and
$f(d_{i}, a_{j})$ represents contributions to artifact
$a_{j}$ by developer $d_{i}$. If a contribution to
this artifact was made by this developer, then the element is 1, otherwise
it is zero. Figure~\ref{fig:network_expl} depicts a situation where multiple developers make commits to multiple functions, some of which are coupled. With respect to Figure~\ref{fig:network_expl}, all of the commit edges
between developers and the functions they contributed to are expressed
in $A_{\text{contrib}}$, the coupling relationships between functions are expressed in another matrix described below.

We compute a matrix that represents the semantic coupling between artifacts
using latent semantic indexing (see Section~\ref{sec:artifact_coupling}) denoted by $A_{\text{coupling}}$. We have chosen this specific approach because it has shown promising results in reflecting developers' perception of software coupling~\citep{bavota2013} (see Section~\ref{sec:ev_background}). Alternative approaches such as Latent Dirichlet Allocation would also be suitable for this purpose~\citep{baeza1999modern}, but they have not yet  shown to produce valid results, as is the case for latent semantic indexing. A detailed description of our approach is contained in the Appendix A. 

We represent the coupling for $N$ artifacts in an $N\times N$ matrix as
\begin{align}
A_{\text{coupling}} &= 
\begin{bmatrix}
\phi(a_{1}, a_{1}) &  \hdots & \phi(a_{1}, a_{N})\\ 
 \vdots & \ddots & \vdots \\ 
\phi(a_{N}, a_{1}) & \hdots & \phi(a_{N}, a_{N})
\end{bmatrix},
\end{align}
where $A_{\text{coupling}}$ is the artifact-coupling matrix and 
$\phi(a_{i}, a_{j})=1$ if the two artifacts $a_{i}$ and
$a_{j}$ are coupled and $\phi(a_{i}, a_{j})=0$ otherwise.
In our study, this matrix represents the semantic coupling
between functions. In reference to Figure~\ref{fig:network_expl}, the
coupling edges between all functions are expressed
in $A_{\text{coupling}}$.

Finally, we combine the information contained in both matrices using the following
operation
\begin{align}\label{d_contrib}
D_{\text{coord}} &= A_{\text{contrib}} \times A_{\text{coupling}} \times A_{\text{contrib}}^\top,
\end{align}
where $D_{\text{coord}}$ is the developer-coordination matrix, with elements that 
represent whether a coordination requirement between two developers exists.
Intuitively, the matrix operation expressed in Equation~\ref{d_contrib} is computing the
developers' mutual dependencies based on contributions to common artifacts and artifacts
that are semantically coupled. The resulting matrix $D_{\text{coord}}$ is symmetric with
respect to the principal diagonal. For our analysis purposes, a weight on certain edges is not
needed and so we assign diagonal elements 0 to prevent the loop edges from influencing the results. 

We will now go through a minimal example to illustrate precisely how this procedure operates.
Suppose that developer $d_{1}$ makes a commit to function $f_{1}$, developer $d_{2}$
makes a commit to function $f_{2}$ and $f_{3}$, and developer $d_{3}$ makes a commit to
function $f_{3}$. Suppose also that the functions $f_{1}$ and $f_{2}$ are
semantically coupled. Figure~\ref{fig:network_expl} illustrates this particular
situation. In our framework, the resulting contribution and coupling matrices are as follows.
\begin{align}
A_{\text{contrib}} = 
\begin{blockarray}{cccc}
& f_{1} & f_{2} & f_{3}\\
\begin{block}{c[ccc]}
d_{1} & 1 & 0 & 0\\ 
d_{2} & 0 & 1 & 1 \\
d_{3} & 0 & 0 & 1 \\
\end{block}
\end{blockarray}
\qquad
A_{\text{coupling}} = 
\begin{blockarray}{cccc}
& f_{1} & f_{2} & f_{3} \\
\begin{block}{c[ccc]}
f_{1} & 1 & 1 & 0\\ 
f_{2} & 1 & 1 & 0\\
f_{3} & 0 & 0 & 1\\
\end{block}
\end{blockarray}
\end{align}
To compute the developer network, we combine the two matrices according to Equation~\ref{d_contrib}.
\begin{align}
D_{\text{coord}} &=
\begin{bmatrix}
1 & 0 & 0\\
0 & 1 & 1\\
0 & 0 & 1\\
\end{bmatrix}
\times
\begin{bmatrix}
1 & 1 & 0\\
1 & 1 & 0\\
0 & 0 & 1\\
\end{bmatrix}
\times
\begin{bmatrix}
1 & 0 & 0\\
0 & 1 & 0\\
0 & 1 & 1\\
\end{bmatrix}
\\
&=
\begin{blockarray}{cccc}
& d_{1} & d_{2} & d_{3}\\
\begin{block}{c[ccc]}
d_{1} & 1 & 1 & 0\\
d_{2} & 1 & 2 & 1\\
d_{3} & 0 & 1 & 1\\
\end{block}
\end{blockarray}
\end{align}
The adjacency matrix correctly expresses an edge between $d_{1}$ and $d_{2}$ because of their commits to coupled functions $f_{1}$ and $f_{2}$. Additionally, an edge exists between developers $d_{2}$ and $d_{3}$ because they both contributed to function $f_{3}$.

\subsection{Developer Network Stream}
\begin{figure}[!t]
	\centering
	\includegraphics[width=0.8\linewidth]{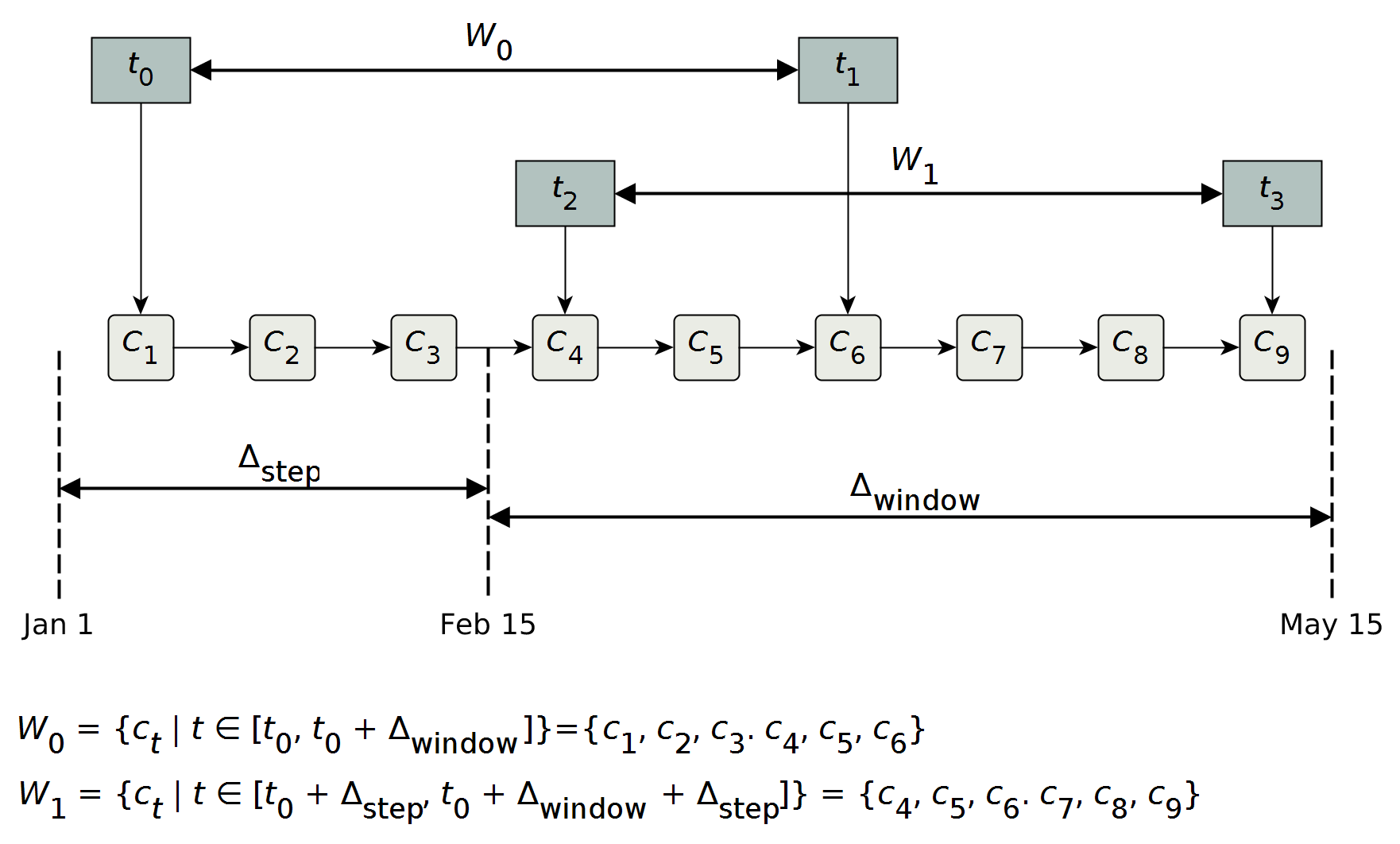}
	\caption{A sequence of commits are shown in chronological order labeled $\{\text{c}_1,\dotsc, \text{c}_9\}$. Two subsequent analysis windows denoted by $\text{W}_0$ and $\text{W}_1$ define which commits are included in each analysis window. The corresponding parameters for $\Delta_{\text{step}}$ and $\Delta_{\text{window}}$ used to define the sliding window process are also shown. Notice that both $\text{W}_0$ and $\text{W}_1$ include $\{\text{c}_4, \text{c}_5,\text{c}_6\}$ so that there is continuity in temporally close activities performed by developers over subsequent analysis windows.}
	\label{fig:vcs_window}
\end{figure}

In a second step, we capture the time-resolved evolution of developer organization by applying a graph-data-stream model to the network-construction procedure; 
a project's history is segmented into sequential overlapping 
observation windows, where each observation window captures a finite range of 
development activity. To linearlize the development history, we flatten the
master branch of the version-control system, which is essentially the
linearization of a directed acyclic graph. All commits are then temporally ordered
using the commit time. The $n^{\text{th}}$ observation window is defined as a set
$W_{n}$ of commits, 
such that $W_{n} = \{commit_{t} \mid t \in [t_{0} + n \cdot \Delta_{\text{step}}, t_{0} + n 
\cdot t_{\text{step}} + \Delta_{\text{window}}]\}$. Where $commit_{t}$ is the commit occurring at time 
$t$, $t_{0}$ is the time of the initial commit, $\Delta_{\text{window}}$ is the window size, and 
$\Delta_{\text{step}}$ is the step size. Figure~\ref{fig:vcs_window} provides a depiction of the sliding-window setup for a linearized history of 9 commits. Since software projects
typically have long-term trends (e.g., number of contributing developers),
the evolution is temporally dependent and must be treated as a nonstationary process.
This implies that the statistics (e.g., mean and variance of the metrics) will vary depending on when the project is observed.
To properly analyze project evolution, we use a small enough observation 
window (90 days) for which the development activity has been shown to be quasi stationary~\citep{Meneely2011}---a technique that is frequently employed in other domains with temporally-dependent processes~\citep{huang1998empirical}. To avoid 
artifacts that arise from aliasing and discontinuities between the edges of the observation windows,
we opted for an overlapping-window 
technique~\citep{huang1998empirical} with a step size that is half of the window size. While smaller step sizes may be
better, because of greater temporal resolution, we observed that using a smaller step size
did not change the results, but 
did significantly add to the computational costs.
In contrast, increasing the step size so that the windows did not overlap obscured 
periodic components in the data.

For each time window, we construct a network to represent the topology of developer 
coordination during a finite time range. The sequence of all finite windows generates
the graph stream capturing the dynamic evolution of developer coordination over the entire
project history. Each graph stream is then processed to extract a multivariate time series
composed of the measurements that quantify the concepts of scale freeness, modularity, hierarchy, in addition to other context features, such as network size.

\subsection{Developer Transitions}\label{sec:markov_chain}
\begin{figure}[!t]
\centering
\normalsize
\begin{tikzpicture}[->,>=stealth',shorten >=1pt,auto,node distance=4cm,
  thick,main node/.style={ellipse,fill=gray!20,draw}]
    
  \node[main node] (1) {Peripheral};
  \node[main node] (2) [below left of=1] {Core};
  \node[main node] (3) [below right of=2] {Isolated};
  \node[main node] (4) [below right of=1] {Absent};

  \path[every node/.style={font=\sffamily\small}]
    (1) edge node [left] {11\%} (4)
        edge [bend right] node[left] {5\%} (2)
        edge [loop above] node {83\%} (1)
        edge [bend right] node[left]{1.7\%} (3)
    (2) edge node [right] {16\%} (1)
        edge [bend left] node {0.5\%} (4)
        edge [loop left] node {84\%} (2)
        edge [bend right] node[left] {0.1\%} (3)
    (3) edge node [right] {0\%} (2)
        edge node [left] {16\%} (4)
        edge [loop below] node {51\%} (3)
        edge [bend right] node[right] {33\%} (1)
    (4) edge [bend left] node {0\%} (2)
        edge [bend right] node[right] {0.8\%} (1)
        edge [bend left] node {0.1\%} (3)
        edge [loop right] node {99.1\%} (4);

\end{tikzpicture}
\caption{The developer-group stability for QEMU shown in the form of a
Markov chain. In some states, the addition of outgoing edge probabilities may not equal unity due to rounding errors.}
\label{fig:markov_chain} 
\end{figure}
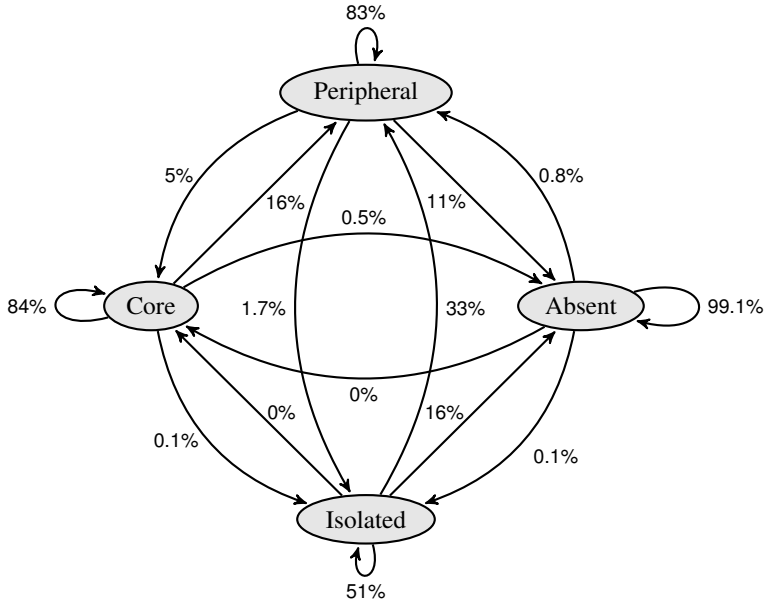

Developer turnover---the process by which developers enter and withdraw from a project---provides important insights into the stability of the organizational structure of a project (see Section~\ref{sec:ev_background}). We define the stability of a group of developers as the probability that members of this group leave
the project by not committing to the version-control system within 90 days of any prior commit. We also expand on this concept by not only studying the likelihood that developers leave a project but also the likelihood of transitioning between different roles in the project. Particularly, we employ sequential-data modeling techniques to formally address this aspect of network evolution. We make use of the discrete state Markov model~\citep{bishop2006pattern} by assigning a discrete state to every developer in the project for each time window. In Appendix B, we discuss the trade-off involved in the choice of analysis windows and the influence it has on the Markov model. In our setting, a developer is able to occupy one of the following 4 possible states.
\begin{itemize}[leftmargin=*]
\item[] \emph{Core}: active developers with degree in the upper 20th percentile
\item[] \emph{Peripheral}: active non-core developers with non-zero degree
\item[] \emph{Isolated}: active developers with a zero degree
\item[] \emph{Absent}: developers that did not make any commits (i.e. no activity)
\end{itemize}
\noindent
We classify developers based on node degree because it has been show to better reflect developer perception than commit count~\citep{joblin2016}. We consider developers with a zero degree to be distinct from the peripheral group since these developers tend to be extremely inactive, often with only a single contribution.
To compute the transition probabilities, each developer's state transitions are expressed by a sequence of random variables $X_{t} \in \{s_{1},s_{2},s_{3},s_{4}\}$ that can take on any of the four states. We then employ the Markov property such that $\text{Pr}(X_{t+1} = x \vert X_{1} = x_{1}, X_{2} = x_{2}, \dots, X_{t} = x_{t}) = \text{Pr}(X_{t+1}=x \vert X_{t} = x_{t})$. The assumption is that, to determine the next state transition, only information about the previous state is required. Using this assumption, we are able to represent developer transitions from state to state as an $N\times N$ transition matrix, in which each element indicates the probability of transitioning from any state in the state space $N$ to any other state during the entire project's evolution. We use a maximum-likelihood estimation to solve for each state transition parameter~\citep{bishop2006pattern}. We experimented with
second order Markov chains---more formally $\text{Pr}(X_{t+1}=x \vert X_{t} = x_{t},  X_{t-1} = x_{t-1})$---to test the validity of our assumptions, but the overall insights do not change and so we only show results for the simpler first order Markov chain.\footnote{The second order Markov chain is more complex by including the random variable $X_{t-1}$ in the model, but the vast majority of variance for our data is explained by the first order Markov chain. We concluded that the increase in model complexity is not justified by the improvement in the model's fit.} Figure~\ref{fig:markov_chain} provides an example developer transition Markov chain: Core developers stay in the core state in the following release with a 84\% probability, transition to the peripheral state with 16\% probability, with 0.1\% probability transition to the isolated state, and with 0.5\% probability to the absent state. All transition probabilities are between 0 and 1, and the sum of all transitions from a single state is equal to 1, to ensure that the conditions for a probability function are maintained.

%Our study includes the notion of core and peripheral developers for two purposes. First, we know from prior empirical studies~\citep{robles09,Crowston05,Terceiro10} that the core developer group is typically more stable than the peripheral group. We can use this a-priori knowledge to test for evidence that the developer network structure correctly captures the notion of core and peripheral developers in the node degree. Second, we can make use of the classification of nodes to provide insightful context for the explanations of the evolutionary adaptions that we observe in the networks' structure.

%\begin{table}[t]
%\caption{Developer Markov chain}
%\centering
%\begin{tabular}{ l | c | c | c | c}
%& Core & Peripheral & Isolated & Absent \\
%\hline Core & 0.9 & 0.1 & 0 & 0 \\
%\hline Peripheral & 0.2 & 0.7 & 0 & 0.1 \\
%\hline Isolated & 0 & 0.25 & 0.25 & 0.5 \\
%\hline Absent & 0 & 0.1 & 0.4 & 0.5 \\
%\hline
%\end{tabular}
%\label{table:markov_chain} 
%\end{table}

\subsection{Network Topology}\label{sec:scale-free}
%For example, approximately 80\% of the internet 
%routers 
%could experience random failures and there would still remain a path between any two 
%routers (i.e., 
%a connected graph). 

%This is because most of the network edges are concentrated in a small 
%number of 
%hub nodes, as long as most of the hubs remain functional, no serious structural damage 
%occurs to 
%the network. The compromise is that a targeted deletion of only a few hubs can isolate 
%large 
%portions of the network.

%Networks that have edges chosen 
%according to 
%a binomial distribution are not resilient to random failures, and through the process of 
%evolution 
%and optimization they have largely been eliminated from many complex network topologies. 

%Secondly, 
%since there is such an unequal concentration of edges, instead of dedicating resources 
%and 
%attention to all nodes equally, there is a disproportionately large benefit to targeting 
%attention 
%on the small number of hub nodes. For example, a scale-free network implies that only the 
%hub nodes 
%need to collaborate effectively and it is largely resilient to collaboration breakdowns 
%in the 
%majority of non-hub nodes. In software engineering resources are often limited and the 
%benefit of a 
%power law is that it represents a significant opportunity to dedicate limited resources 
%in a 
%targeted manner in order to achieve the greatest benefit.

To determine whether a network is scale free, one needs to show that the degree distribution
is explained by a power law (see Section~\ref{sec:ev_background}). There are a number of frequently experienced pitfalls in trying
to test a network for scale freeness, and for this reason, we dedicate significant effort
to ensuring statistical rigor.
A necessary but not sufficient condition for power-law behavior is that the log--log scaled
degree distribution is described by a linear relationship, which makes identifying a power law
involved and challenging~\citep{Clauset2009,goldstein2004}.
Further complicating the situation, it is often the case that empirical data exhibit a power
law only in the tail of the degree distribution, so the power-law model is rarely valid
for the entire set of observations.

In our study, we use a maximum-likelihood technique to solve directly for the power-law
model parameters. Then, we perform a Kolmogorov-Smirnov goodness-of-fit test on the fitted model.
For the moment, let us assume that the lower bound for the power law, $k_{\text{min}}$, is known, then the
power-law scaling parameter, $\alpha$, can be solved using to the following
approximation to the maximum-likelihood estimate:
%
%A more appropriate approach is to characterize the distribution by solving for $\alpha$ using a 
%maximum likelihood approach and a goodness-of-fit test, as we do in this paper. 
%Another important consideration is that it is often the case that empirical data which follows a 
%power law, does so only for the tail region of the distribution. In particular, the power law is 
%valid only beyond a lower bound such that $x > x_{min}$, therefore the distribution is most 
%accurately characterized by solving for both $\alpha$ and $x_{min}$.
%
%We now outline the basic procedure for determining whether a power law is a plausible fit 
%to the observed data, however, we cannot know if other distributions are equally plausible 
%without specifically testing for them. As previously mentioned, using least squares fitting to 
%estimate a linear model to the log-log transformed distribution can have large systematic errors 
%under relatively common conditions~\citep{Clauset2009}, and for this reason we opt for a maximum 
%likelihood approach to estimate the model parameters. The first step is to estimate 
%
\begin{equation} \label{eq:alpha}
\hat{\alpha} \simeq 1 + n { \left [ \sum_{i=1}^{n} \ln \frac{k_{i}}{ k_{\text{min}} - 
\frac{1}{2} } \right 
]}^{-1},
\end{equation}
where $n$ is the number of nodes in the network and $k_{i}$ is the degree of node $i$~\citep{Clauset2009}. 
The choice of $k_{ \text{min}}$ is critical because choosing too low a value will bias
the result by trying to fit a power law to data that do not obey a power law. In contrast, too large a value results in throwing away useful samples for estimating $\alpha$ and will increase the statistical error. 
We solve for the optimal $k_{\text{min}}$ iteratively by selecting a value, solving for $\alpha$ using Equation~\ref{eq:alpha}, and
then testing the fit using the Kolmogorov-Smirnov (KS) statistic. The optimal value for $k_{\text{min}}$ is then
chosen based on the best fit according to the KS statistic.

%One should always keep in mind that if the sample size becomes small, less than 50, the estimates 
%for $\alpha$ become increasingly unreliable~\citep{Clauset2009}. We used an approach proposed by 
%Clauset \emph{et al.}~\citep{clauset2007}, where $x_{min}$ is chosen to be the value which makes the 
%distance, according to the Kolmogorov-Smirnov statistic, between the observed data and the best fit 
%power law, as small as possible for $x > x_{min}$. The $\alpha$ parameter for the power law is 
%estimated using the method of maximum likelihood. Using this approach, one is able to prove that 
%accurate parameter estimates are generated in the limit of large sample size\TODO{cite stats 
%book}. There exist no exact close-form expression for estimating $\alpha$, however we estimated 
%$\alpha$ according to

Once we have solved Equation~\ref{eq:alpha} for the best-fit power law, we still do not know
whether a power law is a plausible model for describing the observed data. For this purpose,
we perform a goodness-of-fit test. This test will discern whether the discrepancy can be explained by
finite random sampling or because the power law is not an appropriate model. To perform the test,
we generate an ensemble of synthetic data sets by sampling the fitted power
law.\footnote{We chose the number of synthetic data sets to generate to introduce 
a precision tolerance of two decimal places in the $p$ value.} Then, we fit a
power law to the synthetic data sets. Next, we use the KS statistic to 
compute the distance between the empirical distribution and its corresponding fitted power law. We 
also compute the KS statistics for each member of the synthetic ensemble and the corresponding fitted power law~\citep{Clauset2009}.
Finally, we compute a $p$ value
that represents the fraction of KS statistics from the synthetic data sets that exceed the
empirical one. A large $p$ value indicates that the deviation of the observed data 
from the fitted model can be attributed to statistical fluctuation and not to a systematic 
error from the selection of an inappropriate model. For $p < 0.05$, we reject the null hypothesis
that the observed data are described by a power law. 

When there are too few samples for statistical tests to be reliable, which is
common early in the project history,
we instead use the Gini coefficient~\citep{atkinson1970} to characterize
the amount of inequality in the network's degree distribution.
The Gini coefficient is bounded between 0 and 1, where 1 indicates strong inequality (i.e., possibly scale free); 0 indicates strong equality (i.e., not scale free).
By definition, scale-free networks contain hub nodes, and as a result, there
is strong inequality in the distribution of edges connecting nodes.
From this, we conclude that a high Gini coefficient is a necessary condition for
scale freeness, and if the network has a low Gini coefficient, it cannot be scale free.

\section{Study \& Results}\label{sec:eval_result}
\subsection{Hypotheses}
We now present and discuss four hypotheses regarding the evolution of developer coordination.
The hypotheses refine our research questions concerning the patterns observed in the evolution of developer networks
and the relationship between the patterns and project scale (see Section~\ref{sec:intro}). The mapping between our research questions and hypotheses is the following:
H1 is to provide better context for interpreting the other hypotheses,
H2 and H4 are related to RQ1 and RQ2,
H3 is related to RQ1.

\vskip 1ex
\noindent\textbf{H1}---\emph{Core developers (those with a node degree in the upper 20th percentile) exhibit significantly greater stability than peripheral developers (those with a node degree in the lower 80th percentile).}
\vskip 1ex
\noindent
When developers withdraw from a project, there are potentially severe consequences as a result of the loss of knowledge and the additional resources required to mentor new developers (see Section~\ref{sec:core_peri}). However, many successful open-source software projects have adapted to benefit from an abundant supply of a group of peripheral
developers that is inherently unstable in comparison to the group of core developers. Since one of the
primary driving forces of change in an organization stems
from developer turnover, it is paramount to understand how an organization structure
may be affected by turnover. We expect that the distinct turnover characteristics
of the groups of core and peripheral developers are responsible for some of the observable
structural features in the developer networks. Understanding the stability patterns of different developer roles can help practitioners to understand the potential risks associated with each role. If stability is uniform across roles, then it may be hard to implement strategies or organizational structures that mitigate risk. However, if stability is substantially greater in one group than in another, it would be sensible to mitigate risk by delegating responsibilities that demand long-term involvement to those individuals that are most likely to be stable. 

\vskip 1ex
\noindent\textbf{H2}---\emph{Long-term sustained growth, in the number of developers contributing to the project, coincides with a scale-free developer coordination structure.}
\vskip 1ex
\noindent
It is almost folklore that adding developers to a project often has the opposite of
the intended outcome, which is that adding developers will accelerate development~\citep{Brooks1975}. 
On this basis, there are important insights that can be gleaned
from observing the changes that occur in the organizational structure as the project grows.
Coordination of a large number of developers demands specialized coordination mechanisms, because 
the number of potential interactions among developers is quadratic in the number of 
developers~\citep{Brooks1975}. Additionally, since the peripheral developer group, representing the majority of open-source software developers, is conjectured to be unstable (see Section~\ref{sec:core_peri}), the implication is that a healthy developer network must be robust to node removals. Therefore, we expect that large developer groups self-organize into scale-free 
networks as an optimization for mitigating the coordination overhead and achieving resilience to coordination 
breakdowns (see Section~\ref{sec:scale_free_networks}). Following the reasoning of \citet{Brooks1975},
as the developer network grows, we expect, at some point, the developer count should
stagnate or decrease, because of ineffective coordination leading to a loss of productivity and 
developers' motivation to participate. It has been shown that, in some situations,
Brooks' law does not apply to open-source software projects~\citep{koch2004}, and we hypothesize that scale freeness
is a reasonable principle to explain this observation.
Therefore, we expect very large projects to exhibit the scale-freeness property as a mechanism to maintaining
productivity despite the potentially enormous coordination costs and risks imposed by a large but unstable peripheral developer group. Finally, scale freeness is an emergent property
of a self-organizing system that is motivated by necessity. Since small developer groups do not benefit from a
scale-free network structure as much as large developer groups, we do not expect small projects
with a small number of developers to form scale-free networks. If scale freeness is required for sustainable project growth, this information helps us to identify healthy growth profiles in large projects. Practitioners can make use of this information to promote policies that encourage the addition of developers to a project in a similar manner as preferential attachment (see Section~\ref{sec:scale_free_networks}) so that the organizational structure reaches a scale-free state.   

\vskip 1ex
\noindent\textbf{H3}---\emph{Developers initially form loosely connected groups that are not internally well connected (i.e., that have a low modularity). As time proceeds, developer groups tend to become more strongly connected in terms of the clustering coefficient until an upper bound is reached.}
\vskip 1ex
\noindent
As a project evolves, several factors encourage developers to coordinate, but there are
also opposing forces. Based on prior experience and empirical evidence,
software evolution tends to cause an increase along several project dimensions 
(e.g., lines of code, complexity, number of developers, etc.) and
will demand increasing levels of coordination between developers to avoid system degradation~\citep{lehman1997metrics,lehman2001}. 
Furthermore, it is reasonable to expect that developers will become more familiar with each other and rely on the knowledge of others for support in the completion of development tasks. 
Empirical evidence from studies on various open-source software projects also suggests that
developers tend to specialize on particular artifacts (e.g., subsystems or files)
and form groups with common responsibilities and shared mental models~\citep{koch2004,joblin2015}.
These influences increase modularity in the developer network by causing additional edges to form in
local sub-networks that are dedicated to a particular responsibility. 
The opposing force arises from the quadratic scaling between the number of developers and potential
coordination relationships, where the cost of coordination can easily dominate the benefit
achieved from coordination~\citep{Brooks1975}.
Therefore, developer coordination is constrained to evolve in a manner that balances these
opposing forces. We expect that an equilibrium exists between the benefit and cost of coordination,
and this will govern the evolution of developer coordination. If this hypothesis is confirmed, it suggests that the socio-technical environment in which developers work changes substantially over time. As a result, there are presumably changes in the coordination challenges that developers face and they would likely benefit from different techniques and tools to support coordination during different phases of project.

\vskip 1ex
\noindent\textbf{H4}---\emph{In early project phases, the developer-coordination structure is
hierarchically arranged. As a project grows and matures, the developer-coordination structure
will gradually converge to a network that does not exhibit hierarchy, as the command-and-control
structure becomes more distributed.}
\vskip 1ex
\noindent
A project's command-and-control structure is responsible for directing the work of others in a coordinated manner. 
In the early phases of a project, it is conceivable that the small number of initial developers have a comprehensive
understanding of the global project details and are capable of effectively coordinating the work with others in a centralized configuration.
In these early project conditions, hierarchy is an effective organizational structure because it promotes efficiency through regularity and is appropriate when the developer network is stable~\citep{kotter2014}. As the project evolves and grows in the number of developers and system size,
developer coordination becomes increasingly formidable, especially, once the peripheral developer group has grown to be significantly larger than the core developer group. Empirical evidence indicates that
efficiency in large open-source software projects is the result of self-organizing cooperative and highly decentralized work~\citep{koch2004},
which becomes increasingly important as a project grows. The result is that the command-and-control structure
must evolve to become more distributed, because no single person could reasonably have a comprehensive understanding of
the global project state, and distributed self-organization must take over.
Furthermore, hierarchy is an intrinsically inflexible organizational structure that strongly promotes regularity~\citep{kotter2014},
but as the project evolves, organizational flexibility becomes increasingly important so that
the project can avoid the detrimental misalignment of organizational structure
and the technical structure as a result of evolution~\citep{Sosa2004}. For practitioners, it is often unclear which organizational structures are suitable for different project conditions. It may be the case that different organizational structures are more appropriate during early project conditions and others during late project conditions. By addressing this hypothesis, we gather evidence of how successful open-source software projects evolve with respect to hierarchy. 

\subsection{Subject Projects}
For the purpose of our study, we selected 18 open-source software projects as listed in Table~\ref{table:oss_project_data}.
The subject projects vary in the following dimensions: (a) size (source lines of code,
from 50\,KLOC to over 16\,MLOC, number of developers from 25 to 1000), (b) age
(days since first commit), (c) technology (programming language, libraries),
(d) application domain (operating system, development, productivity,
etc.), and (e) version-control system used (Git, Subversion).
We chose these projects because they are all widely
deployed, and have long development histories.
The data and list of figures for all projects are available at
the supplementary Web site.

\begin{table*}[t!]
  \begin{threeparttable}
  \centering
  \caption{Overview of subject projects}
  \label{table:oss_project_data}
  \begin{tabular}{llllrrrrr}
    \toprule
    \multicolumn{6}{c}{} & \multicolumn{3}{c}{Developer Count} \\
    \cmidrule(r){7-9}
    % latex table generated in R 3.2.2 by xtable 1.7-4 package
% Thu Apr 14 15:27:20 2016
Project & Domain & Lang & Period & SLOC & Commits & Cur. & Max & Min \\ 
  \midrule
Apache HTTP & Server & C & 05/99--06/15 & 2M & 73K & 13 & 26 &  2 \\ 
  Chromium & User & C/++, JS & 07/08--06/15 & 16M & 533K & 642 & 1056 & 71 \\ 
  Django & Devel & Python & 07/05--01/15 & 400K & 38K & 98 & 105 &  3 \\ 
  Firefox & User & C/++, JS & 03/98--06/15 & 12M & 230K & 417 & 474 & 62 \\ 
  GCC & Devel & C/++ & 06/91--01/15 & 7M & 137K & 117 & 122 &  2 \\ 
  Homebrew & User & Ruby & 05/09--06/15 & 100K & 42K & 473 & 525 &  3 \\ 
  Joomla & CMS & PHP & 09/05--06/15 & 400K & 20K & 53 & 78 &  2 \\ 
  jQuery & Devel & JS & 03/06--06/15 & 65K & 12K &  5 & 30 &  2 \\ 
  Linux & OS & C & 04/05--05/15 & 17M & 570K & 1445 & 1512 & 481 \\ 
  LLVM & Devel & C/++ & 06/01--06/15 & 1.2M & 120K & 127 & 128 &  3 \\ 
  Mongo & Database & C/++, JS & 10/07--06/15 & 600K & 28K & 45 & 53 &  2 \\ 
  Node.js & Devel & C/++, JS & 04/09--05/15 & 5M & 23K & 19 & 53 &  2 \\ 
  PHP & Devel & PHP, C & 04/99--05/15 & 2.5M & 100K & 46 & 66 &  9 \\ 
  QEMU & OS & C & 11/05--06/15 & 1M & 37K & 116 & 157 &  2 \\ 
  Qt 4 & Devel & C++ & 03/09--04/15 & 1.5M & 36K &  7 & 122 &  5 \\ 
  Rails & Devel & Ruby & 11/04--06/15 & 200K & 49K & 146 & 213 &  2 \\ 
  Salt & Devel & Python & 02/11--06/15 & 200K & 44K & 204 & 205 &  3 \\ 
  U-Boot & Devel & C & 12/02--06/15 & 1.2M & 32K & 114 & 134 &  2 \\ 
   \bottomrule

  \end{tabular}
  \end{threeparttable}
\end{table*}

\subsection{Developer-Group Stability}
To address H1, we now present the results regarding the stability of the groups of core and peripheral developers. We applied the procedure described in Section~\ref{sec:markov_chain} to construct a Markov chain representing the transitions between the four possible developer states (core, peripheral, isolated, and absent). Considering the evolution of all subject projects, the primary finding is that core developers are significantly less likely to withdraw compared to non-core developers. In Table~\ref{table:project_markov_chain}, we
present the Markov chain for each subject project.
\begin{table}
\caption{Probabilities of developer state transitions}
{\tabcolsep=0.07cm
\begin{subtable}{0.5\textwidth}
\centering
% latex table generated in R 3.2.2 by xtable 1.7-4 package
% Thu Apr 14 15:27:01 2016
\begin{tabular}{rrrrr}
  \toprule
 & absent & core & isolated & peripheral \\ 
  \midrule
absent & 0.964 & 0.003 & 0.007 & 0.027 \\ 
  core & 0.017 & 0.617 & 0.010 & 0.356 \\ 
  isolated & 0.322 & 0.022 & 0.291 & 0.365 \\ 
  peripheral & 0.181 & 0.103 & 0.077 & 0.639 \\ 
   \bottomrule
\end{tabular}
\caption{Apache HTTP}
\end{subtable}
\begin{subtable}{0.5\textwidth}
\centering
% latex table generated in R 3.2.2 by xtable 1.7-4 package
% Thu Apr 14 15:27:01 2016
\begin{tabular}{rrrrr}
  \toprule
 & absent & core & isolated & peripheral \\ 
  \midrule
absent & 0.978 & 0.001 & 0.000 & 0.021 \\ 
  core & 0.017 & 0.698 & 0.000 & 0.285 \\ 
  isolated & 0.347 & 0.006 & 0.136 & 0.511 \\ 
  peripheral & 0.170 & 0.089 & 0.003 & 0.738 \\ 
   \bottomrule
\end{tabular}
\caption{Chromium}
\end{subtable}
\begin{subtable}{0.5\textwidth}
\centering
% latex table generated in R 3.2.2 by xtable 1.7-4 package
% Thu Apr 14 15:27:01 2016
\begin{tabular}{rrrrr}
  \toprule
 & absent & core & isolated & peripheral \\ 
  \midrule
absent & 0.982 & 0.002 & 0.002 & 0.015 \\ 
  core & 0.095 & 0.680 & 0.006 & 0.218 \\ 
  isolated & 0.433 & 0.022 & 0.239 & 0.306 \\ 
  peripheral & 0.332 & 0.081 & 0.027 & 0.559 \\ 
   \bottomrule
\end{tabular}
\caption{Django}
\end{subtable}
\begin{subtable}{0.5\textwidth}
\centering
% latex table generated in R 3.2.2 by xtable 1.7-4 package
% Thu Apr 14 15:27:01 2016
\begin{tabular}{rrrrr}
  \toprule
 & absent & core & isolated & peripheral \\ 
  \midrule
absent & 0.987 & 0.000 & 0.001 & 0.012 \\ 
  core & 0.012 & 0.700 & 0.001 & 0.287 \\ 
  isolated & 0.349 & 0.007 & 0.208 & 0.436 \\ 
  peripheral & 0.193 & 0.092 & 0.017 & 0.698 \\ 
   \bottomrule
\end{tabular}
\caption{Firefox}
\end{subtable}
\begin{subtable}{0.5\textwidth}
\centering
% latex table generated in R 3.2.2 by xtable 1.7-4 package
% Thu Apr 14 15:27:01 2016
\begin{tabular}{rrrrr}
  \toprule
 & absent & core & isolated & peripheral \\ 
  \midrule
absent & 0.992 & 0.000 & 0.001 & 0.007 \\ 
  core & 0.000 & 0.823 & 0.000 & 0.177 \\ 
  isolated & 0.155 & 0.001 & 0.556 & 0.288 \\ 
  peripheral & 0.057 & 0.050 & 0.017 & 0.876 \\ 
   \bottomrule
\end{tabular}
\caption{GCC}
\end{subtable}
\begin{subtable}{0.5\textwidth}
\centering
% latex table generated in R 3.2.2 by xtable 1.7-4 package
% Thu Apr 14 15:27:01 2016
\begin{tabular}{rrrrr}
  \toprule
 & absent & core & isolated & peripheral \\ 
  \midrule
absent & 0.969 & 0.006 & 0.000 & 0.026 \\ 
  core & 0.246 & 0.500 & 0.000 & 0.253 \\ 
  isolated & 0.500 & 0.000 & 0.500 & 0.000 \\ 
  peripheral & 0.433 & 0.095 & 0.000 & 0.473 \\ 
   \bottomrule
\end{tabular}
\caption{Homebrew}
\end{subtable}
\begin{subtable}{0.5\textwidth}
\centering
% latex table generated in R 3.2.2 by xtable 1.7-4 package
% Thu Apr 14 15:27:01 2016
\begin{tabular}{rrrrr}
  \toprule
 & absent & core & isolated & peripheral \\ 
  \midrule
absent & 0.979 & 0.002 & 0.004 & 0.016 \\ 
  core & 0.107 & 0.638 & 0.005 & 0.251 \\ 
  isolated & 0.416 & 0.032 & 0.348 & 0.204 \\ 
  peripheral & 0.298 & 0.095 & 0.051 & 0.556 \\ 
   \bottomrule
\end{tabular}
\caption{Joomla}
\end{subtable}
\begin{subtable}{0.5\textwidth}
\centering
% latex table generated in R 3.2.2 by xtable 1.7-4 package
% Thu Apr 14 15:27:01 2016
\begin{tabular}{rrrrr}
  \toprule
 & absent & core & isolated & peripheral \\ 
  \midrule
absent & 0.980 & 0.001 & 0.001 & 0.018 \\ 
  core & 0.051 & 0.624 & 0.006 & 0.318 \\ 
  isolated & 0.250 & 0.125 & 0.000 & 0.625 \\ 
  peripheral & 0.344 & 0.088 & 0.004 & 0.564 \\ 
   \bottomrule
\end{tabular}
\caption{jQuery}
\end{subtable}
\begin{subtable}{0.5\textwidth}
\centering
% latex table generated in R 3.2.2 by xtable 1.7-4 package
% Thu Apr 14 15:27:01 2016
\begin{tabular}{rrrrr}
  \toprule
 & absent & core & isolated & peripheral \\ 
  \midrule
absent & 0.972 & 0.002 & 0.001 & 0.025 \\ 
  core & 0.053 & 0.681 & 0.001 & 0.264 \\ 
  isolated & 0.401 & 0.016 & 0.184 & 0.399 \\ 
  peripheral & 0.291 & 0.086 & 0.009 & 0.614 \\ 
   \bottomrule
\end{tabular}
\caption{Linux}
\end{subtable}
\begin{subtable}{0.5\textwidth}
\centering
% latex table generated in R 3.2.2 by xtable 1.7-4 package
% Thu Apr 14 15:27:01 2016
\begin{tabular}{rrrrr}
  \toprule
 & absent & core & isolated & peripheral \\ 
  \midrule
absent & 0.980 & 0.000 & 0.003 & 0.017 \\ 
  core & 0.014 & 0.719 & 0.002 & 0.265 \\ 
  isolated & 0.351 & 0.015 & 0.224 & 0.409 \\ 
  peripheral & 0.187 & 0.093 & 0.029 & 0.690 \\ 
   \bottomrule
\end{tabular}
\caption{LLVM}
\end{subtable}
\begin{subtable}{0.5\textwidth}
\centering
% latex table generated in R 3.2.2 by xtable 1.7-4 package
% Thu Apr 14 15:27:01 2016
\begin{tabular}{rrrrr}
  \toprule
 & absent & core & isolated & peripheral \\ 
  \midrule
absent & 0.974 & 0.000 & 0.004 & 0.022 \\ 
  core & 0.000 & 0.691 & 0.000 & 0.309 \\ 
  isolated & 0.412 & 0.000 & 0.329 & 0.259 \\ 
  peripheral & 0.218 & 0.102 & 0.021 & 0.660 \\ 
   \bottomrule
\end{tabular}
\caption{Mongo}
\end{subtable}
\begin{subtable}{0.5\textwidth}
\centering
% latex table generated in R 3.2.2 by xtable 1.7-4 package
% Thu Apr 14 15:27:01 2016
\begin{tabular}{rrrrr}
  \toprule
 & absent & core & isolated & peripheral \\ 
  \midrule
absent & 0.972 & 0.002 & 0.003 & 0.023 \\ 
  core & 0.119 & 0.686 & 0.008 & 0.187 \\ 
  isolated & 0.538 & 0.015 & 0.288 & 0.159 \\ 
  peripheral & 0.406 & 0.081 & 0.046 & 0.467 \\ 
   \bottomrule
\end{tabular}
\caption{Node.js}
\end{subtable}
\begin{subtable}{0.5\textwidth}
\centering
% latex table generated in R 3.2.2 by xtable 1.7-4 package
% Thu Apr 14 15:27:01 2016
\begin{tabular}{rrrrr}
  \toprule
 & absent & core & isolated & peripheral \\ 
  \midrule
absent & 0.982 & 0.001 & 0.004 & 0.014 \\ 
  core & 0.021 & 0.672 & 0.008 & 0.299 \\ 
  isolated & 0.353 & 0.022 & 0.323 & 0.302 \\ 
  peripheral & 0.198 & 0.099 & 0.059 & 0.643 \\ 
   \bottomrule
\end{tabular}
\caption{PHP}
\end{subtable}
\begin{subtable}{0.5\textwidth}
\centering
% latex table generated in R 3.2.2 by xtable 1.7-4 package
% Thu Apr 14 15:27:01 2016
\begin{tabular}{rrrrr}
  \toprule
 & absent & core & isolated & peripheral \\ 
  \midrule
absent & 0.991 & 0.000 & 0.001 & 0.008 \\ 
  core & 0.005 & 0.837 & 0.001 & 0.158 \\ 
  isolated & 0.164 & 0.000 & 0.507 & 0.330 \\ 
  peripheral & 0.109 & 0.047 & 0.017 & 0.828 \\ 
   \bottomrule
\end{tabular}
\caption{QEMU}
\end{subtable}}
\label{table:project_markov_chain}
\end{table}
\addtocounter{table}{-1}
\clearpage
\begin{table}
{\tabcolsep=0.07cm
\begin{subtable}{0.5\textwidth}
\centering
% latex table generated in R 3.2.2 by xtable 1.7-4 package
% Thu Apr 14 15:27:01 2016
\begin{tabular}{rrrrr}
  \toprule
 & absent & core & isolated & peripheral \\ 
  \midrule
absent & 0.964 & 0.003 & 0.010 & 0.022 \\ 
  core & 0.103 & 0.564 & 0.030 & 0.303 \\ 
  isolated & 0.395 & 0.048 & 0.349 & 0.208 \\ 
  peripheral & 0.257 & 0.101 & 0.055 & 0.588 \\ 
   \bottomrule
\end{tabular}
\caption{Qt 4}
\end{subtable}
\begin{subtable}{0.5\textwidth}
\centering
% latex table generated in R 3.2.2 by xtable 1.7-4 package
% Thu Apr 14 15:27:01 2016
\begin{tabular}{rrrrr}
  \toprule
 & absent & core & isolated & peripheral \\ 
  \midrule
absent & 0.994 & 0.000 & 0.000 & 0.005 \\ 
  core & 0.033 & 0.813 & 0.000 & 0.153 \\ 
  isolated & 0.185 & 0.005 & 0.439 & 0.372 \\ 
  peripheral & 0.147 & 0.045 & 0.010 & 0.798 \\ 
   \bottomrule
\end{tabular}
\caption{Rails}
\end{subtable}
\begin{subtable}{0.5\textwidth}
\centering
% latex table generated in R 3.2.2 by xtable 1.7-4 package
% Thu Apr 14 15:27:01 2016
\begin{tabular}{rrrrr}
  \toprule
 & absent & core & isolated & peripheral \\ 
  \midrule
absent & 0.986 & 0.001 & 0.001 & 0.013 \\ 
  core & 0.020 & 0.834 & 0.001 & 0.146 \\ 
  isolated & 0.162 & 0.003 & 0.545 & 0.290 \\ 
  peripheral & 0.137 & 0.047 & 0.014 & 0.803 \\ 
   \bottomrule
\end{tabular}
\caption{Salt}
\end{subtable}
\begin{subtable}{0.5\textwidth}
\centering
% latex table generated in R 3.2.2 by xtable 1.7-4 package
% Thu Apr 14 15:27:01 2016
\begin{tabular}{rrrrr}
  \toprule
 & absent & core & isolated & peripheral \\ 
  \midrule
absent & 0.980 & 0.003 & 0.001 & 0.016 \\ 
  core & 0.121 & 0.589 & 0.006 & 0.283 \\ 
  isolated & 0.440 & 0.027 & 0.167 & 0.365 \\ 
  peripheral & 0.341 & 0.096 & 0.030 & 0.533 \\ 
   \bottomrule
\end{tabular}
\caption{U-Boot}
\end{subtable}
}
\end{table}

For illustration, we describe the Markov chain for QEMU as a representative of the primary result. Reading across the row labeled ``core'' of the table for QEMU, we see that core developers are very unlikely to leave the project with only a 0.5\% chance. In comparison to developers in the peripheral state and isolated state, the chance of becoming absent is 11\% and 16\% respectively. We see that it is a common phenomenon that the core developers are 5 to 10 times less likely to leave a project in comparison to peripheral developers. This result is convincing evidence that the groups of peripheral and isolated developers, or more generally non-core developers, are significantly less stable than core developers. Furthermore, we see that, once a developer enters the absent state, there is an overwhelming probability that the developer will not return to the project. This result suggests, once a developer becomes absent for a single revision, in most cases, she will not participate in contributing code in the future. Since entering the absent state most likely indicates a total lost of the individual and any valuable knowledge they possess, the peripheral developers introduce risk through their volatility.
\vskip 1ex
\noindent\fbox{\parbox{\linewidth}{\ For all subject projects, our data indicate that the core developer group is
significantly more stable than the peripheral developer group,
and on this basis, we \emph{accept H1}.}}

\subsection{Scale Freeness}
We now discuss the results of applying the procedure described in 
Section~\ref{sec:scale-free} to address H2. The primary goal is to determine whether
a power-law degree distribution is a plausible model for describing the observed developer networks
and thus can be characterized as scale-free networks. We must eliminate 
Apache HTTP from this evaluation because the project has too few developers, and the 
statistical error with small sample sizes can lead to inaccurate conclusions~\citep{Clauset2009}. For the remaining 17 projects, if the goodness-of-fit $p$ value is greater than 0.05, we can confidently conclude that the network is scale free.

\begin{figure*}[t!]
\centering
	\begin{subfigure}{0.49\linewidth}
		\includegraphics[width=\linewidth]{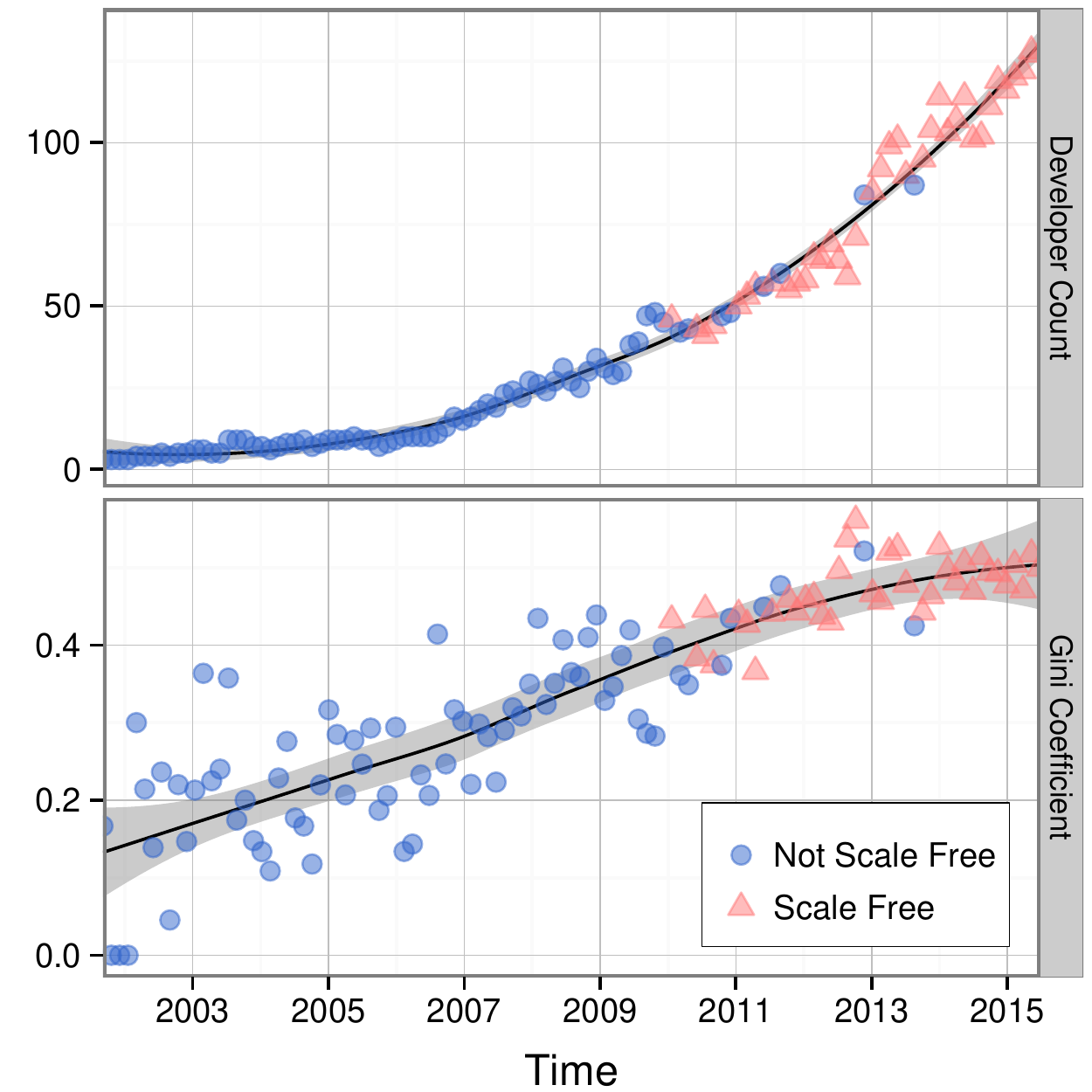}
		\caption{LLVM}
	\end{subfigure}
	\begin{subfigure}{0.49\linewidth}
		\includegraphics[width=\linewidth]{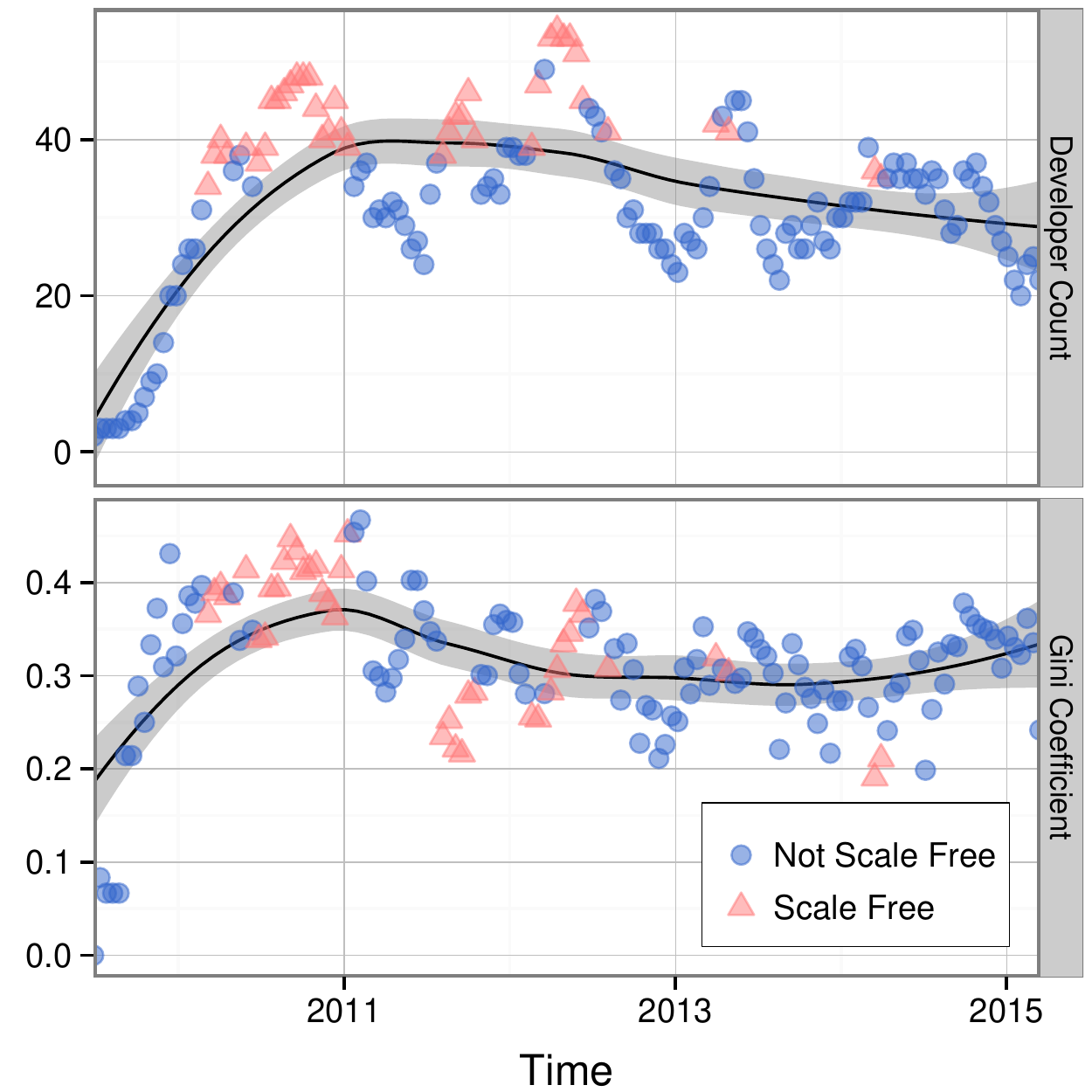}
		\caption{Node.js}
	\end{subfigure}
\caption{Evolutionary profile for entire history of LLVM and Node.js. Time series are shown for the Gini coefficient (bottom) and the number of developers (top). A smooth curve is fitted to the observations with the 99\% confidence interval shown in gray. The shape of the data points indicates
whether the network was scale free for a given point in time.}
\label{fig:growth_profile}
\end{figure*}

One primary finding, which is true for 17 subject projects, is that 
the scale-freeness property is temporally dependent, which means that this property is not universally present with respect to time. This is notable because it is a distinctly different view point from prior studies that approached the topic of scale freeness from a temporally static perspective~\citep{Lopez-fernandez2006}.
To illustrate this result, a typical chronological profile
is shown in the left portion of Figure~\ref{fig:growth_profile}, taken from LLVM. The top figure illustrates the network growth in terms of the number of developers contributing to the project, and the bottom figure illustrates the Gini coefficient. Each sample point represents a measurement for a single developer network that is computed for a single development time window. The shape of the sample point represents whether the network is scale free during the given development window. To help draw attention to the general trends in the data, a smooth curve has been fitted using locally weighted scatterplot smoothing, with 99\% confidence intervals in gray. 

The evolutionary profile of a project is typically composed of the following three distinct temporal phases.
The initial phase, which can last for a number of years, is characterized by extremely limited growth in the number of contributing developers.
In Figure~\ref{fig:growth_profile}, for LLVM, this phase occurs from the project's beginning in 2002 until 2006. 
During this period, the network exhibits high levels of coordination equality, because most developers are similar with respect to the degree of coordination with other developers and so the coordination requirements are uniformly distributed among all developers. The magnitude of 
coordination equality in the network is quantified using the Gini coefficient of the corresponding
degree distribution. This is shown in Figure~\ref{fig:growth_profile}, for LLVM, where we 
see an initially low Gini coefficient (i.e., high equality). A low Gini coefficient indicates that, during this initial phase, 
most developers are similar in their degree of coordination with others, and the network 
is not scale free (i.e., it lacks the characteristic hub nodes discussed in 
Section~\ref{sec:scale-free}).

In the second phase, we see that projects reach a critical mass point, at which a positive 
trend component is visible in the number of contributing developers.
For LLVM (see Figure~\ref{fig:growth_profile}),
this transition point occurred in late 2006 to early 2007 and the second phase lasts until mid 2011.
Following the transition point, super-linear growth with an
increasing slope occurs until the end of the analysis period. During that time, the 
Gini coefficient also has a positive trend component, indicating that the network has progressively less equality, 
because hub nodes, with significantly more coordination requirements than the average developer, begin to form.
During this phase, the scale-freeness property emerges for the first time, but the state of being scale free is
initially unstable. Most of the projects become scale free once the network size has reached roughly 50 developers, but in no case
does a network exceed a size of 86 developers without first becoming scale free. The only exception is for Linux because
we did not have sufficient data to observe the early phases. The number of developers contributing to the project when the first appearances of scale freeness occurs is shown in Table~\ref{table:project_measurements} for all projects under the
column label ``SF Size''. Overall, the two important results from this phase are that,
during time periods with much less than 50 developers, developer equality is relatively high and scale freeness is not a common property.
In contrast, during periods that significantly exceed 50 developers, developer networks are predominantly scale free. Essentially, the scale-freeness property appears to be dependent on the network size and the time of observation.

In 12 projects (Chromium, Django, Firefox, GCC, Homebrew, Joomla, Linux, LLVM, QEMU, Rails, Salt, and U-Boot), a third phase is visible in which the scale-freeness property stabilizes and is rarely lost. 
In Figure~\ref{fig:growth_profile} for LLVM, this phase begins shortly after mid 2011 and extends until the end of the analysis period.
By means of a visual inspection of the time-series data, a number of patterns is clearly visible.
In all of the projects that achieve stable scale freeness (i.e., a scale-free state that is maintained over several consecutive analysis windows), they demonstrate the capability of long-term sustained (e.g., over the period of several years), and often accelerating, growth in the number of contributing developers. This is visible in Figure~\ref{fig:growth_profile} for LLVM, where we observe an increasing slope in the number of developers
year over year, and growth continues on until the end of the analysis period.
In the other 6 projects (Apache HTTP, jQuery, Mongo, Node.js, Qt 4, and PHP), scale freeness is either never achieved, remains unstable 
indefinitely, or is lost indefinitely. The growth profile in these 6 projects is very different from the projects that achieve stable scale freeness, and growth appears to be unsustainable because project growth rates decrease with time and often the number of contributing developers even drops. An example of these two distinct cases is presented in Figure~\ref{fig:growth_profile},
where LLVM reaches stable scale freeness and has long term accelerating growth. In contrast, Node.js does not achieve stable scale freeness and has unsustainable growth with long-term loss of developers. The percentage of time each project spends in a scale-free state is shown in Table~\ref{table:project_measurements} under column ``\% SF''. The measurements indicate that the large projects that have had long term growth spend a significantly larger percentage of time in a scale-free state in comparison to projects without long-term sustained growth.

\begin{figure}[t]
\centering
\includegraphics[width=0.8\linewidth]{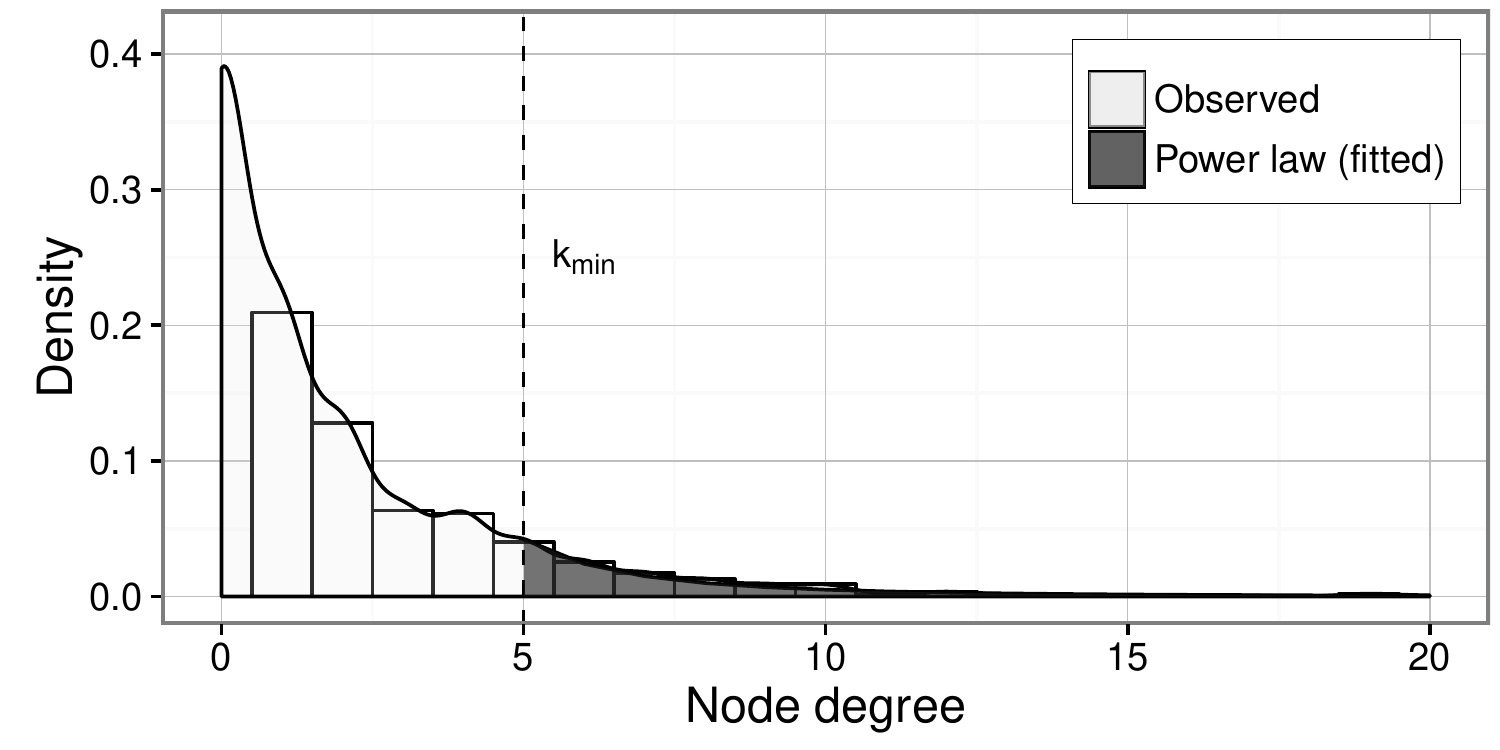}
\caption{Power law fitted to the degree distribution for Linux. The power-law distribution describes the developers with a degree greater than $k_{\text{min}}$, the majority of low degree developers are not described by a power law.}
\label{fig:degree_dist}
\end{figure}

A scale-free network is characterized based on whether the \emph{tail} of the degree 
distribution is described by a power law (see Section~\ref{sec:scale-free}).
A rarely addressed yet important factor, though, is the proportion of nodes that are described by 
the power law. We illustrate
this point in Figure~\ref{fig:degree_dist}, where the tail region described by the power 
law excludes the majority
of developers of the Linux kernel. We found that, in all projects, the proportion of developers 
that are described by the
power law is low and typically ranges between 20\%--50\%. 
We saw that, there is an inverse relationship between the network size and
the percentage of developers that are characterized by a power law. We illustrate the
results for all projects in Table~\ref{table:project_measurements}, where column
``$k_{\text{min}}$'' represents the lower bound for the power-law distribution, column
``\% Dev SF'' represents the percentage of developers the are described by the
power-law distribution, and column ``SF Dev'' represents the absolute number
of developers described by the power-law distribution. These measurements are taken 
from the most recent analysis period, and the corresponding analysis windows are provided
in Table~\ref{table:project_measurements}.

We found two interesting outlier projects with respect to the presence of the scale-freeness 
property. 
For PHP and jQuery, the network size reached a peak at roughly 50 developers for 
several months
yet never reached a stable scale-free state. After the peak, both projects
experienced years of continuous 
loss of developers and never recovered. Another interesting outlier is Firefox, 
where during 
the period of September 2009 to December 2009, the network was frequently and unexpectedly not scale free. While
the definitive cause of the disruption is unknown, we learned that, during 
this period, 
Firefox experienced severe release problems resulting in a major revision being released 
one year late.\footnote{\url{http://www.cnet.com/news/mozilla-pushes-back-firefox-3-6-4-0-deadlines/}}
It is interesting to note that the network-structure disturbances were observable several 
months 
before the first public announcement of release problems.
\vskip 1ex
\noindent\fbox{\parbox{\linewidth}{\ In summary, our study suggests that open-source software projects lack a scale-free structure in the initial 
phases while the developer network scale is small. In developer networks where growth significantly exceeds 50 developers, we 
always observe the emergence of scale freeness and no project ever grew beyond 86 developers
without first becoming scale free. The caveat is that the scale-freeness property is 
temporally dependent, and in some projects, remains in an oscillatory state indefinitely.
Overall, we \emph{accept H2}.}}

\begin{table*}[t!]
  \begin{threeparttable}
  \centering
  \caption{Developer network structural measurements} 
  \label{table:project_measurements}
  \begin{tabular}{l|rrrrrrrrr}
    \toprule
    \multicolumn{1}{c}{} & \multicolumn{5}{c}{Scale Free} &
    \multicolumn{2}{|c|}{Modularity} & \multicolumn{2}{c}{Hierarchy}\\
    \cmidrule(r){2-10}
    % latex table generated in R 3.2.2 by xtable 1.7-4 package
% Thu Apr 14 15:27:20 2016
Project & S.F. size & \%S.F. & $k_{min}$ & \% Dev S.F. & \# Dev S.F. & $\mu_{cc}$ & $\sigma_{cc}^{2}$ & $\beta_{1_{early}}$ & $\beta_{1_{late}}$ \\ 
  \midrule
Apache HTTP & N/A & 0.00 &   7 & 0.46 &   6 & 0.49 & 0.12 & -1.14 & -0.85 \\ 
  Chromium &  71 & 97.60 & 1012 & 0.08 &  53 & 0.44 & 0.04 & -0.40 & -0.35 \\ 
  Django &  42 & 25.00 &  49 & 0.60 &  59 & 0.57 & 0.04 & -2.32 & -0.51 \\ 
  Firefox &  73 & 90.60 & 481 & 0.07 &  30 & 0.48 & 0.05 & -0.43 & -0.27 \\ 
  GCC &  36 & 55.90 & 164 & 0.26 &  30 & 0.43 & 0.04 & -0.73 & -0.40 \\ 
  Homebrew &  80 & 87.50 & 355 & 0.49 & 230 & 0.61 & 0.02 & -1.28 & -0.29 \\ 
  Joomla &  55 & 20.00 &   2 & 0.66 &  35 & 0.38 & 0.16 & -1.67 & -0.66 \\ 
  jQuery &  30 & 1.41 &   5 & 0.80 &   4 & 0.47 & 0.10 & -2.09 & -1.66 \\ 
  Linux & 515 & 98.80 & 1521 & 0.05 &  69 & 0.58 & 0.05 & -0.28 & -0.25 \\ 
  LLVM &  46 & 32.70 &  51 & 0.24 &  30 & 0.45 & 0.08 & -2.22 & -0.39 \\ 
  Mongo &  51 & 3.28 &  30 & 0.67 &  30 & 0.38 & 0.07 & -2.30 & -0.53 \\ 
  Node.js &  35 & 16.30 &   1 & 0.47 &   9 & 0.16 & 0.10 & -2.30 & -1.33 \\ 
  PHP &  46 & 14.60 &  29 & 0.65 &  30 & 0.45 & 0.08 & -0.85 & -0.44 \\ 
  QEMU &  37 & 61.00 &  88 & 0.27 &  31 & 0.53 & 0.05 & -2.19 & -0.35 \\ 
  Qt 4 &  86 & 43.80 &   3 & 0.86 &   6 & 0.68 & 0.12 & -0.50 & -0.92 \\ 
  Rails &  38 & 69.20 &  55 & 0.21 &  30 & 0.58 & 0.08 & -2.17 & -0.34 \\ 
  Salt &  32 & 82.20 &  89 & 0.36 &  74 & 0.55 & 0.07 & -0.95 & -0.31 \\ 
  U-Boot &  41 & 64.60 &  41 & 0.58 &  66 & 0.60 & 0.05 & -1.39 & -0.29 \\ 
   \bottomrule

  \end{tabular}
  \begin{tablenotes}
  \small
  \item S.F. size -- network size at first appearance of scale freeness
  \item \% S.F. -- percent of time the project exhibits scale freeness
  \item $k_{\text{min}}$ -- minimum bound on the power-law distribution
  \item \% Dev S.F. -- percent of developers described by the power-law distribution
  \item \# Dev S.F. -- number of developers described by the power-law distribution
  \item $\mu_{cc}$ -- mean value of clustering coefficient for latest development cycle
  \item $\sigma_{cc}^{2}$ -- variance of clustering coefficient for latest development cycle
  \item ${\beta_{1}}_{early}$ -- slope parameter for an early development cycle
  \item ${\beta_{1}}_{late}$ -- slope parameter for the most recent development cycle
  \end{tablenotes}
  \end{threeparttable}
\end{table*}

\subsection{Modularity}\label{sec:results_modularity}
\begin{figure}[!t]
\centering
\includegraphics[width=0.8\linewidth]{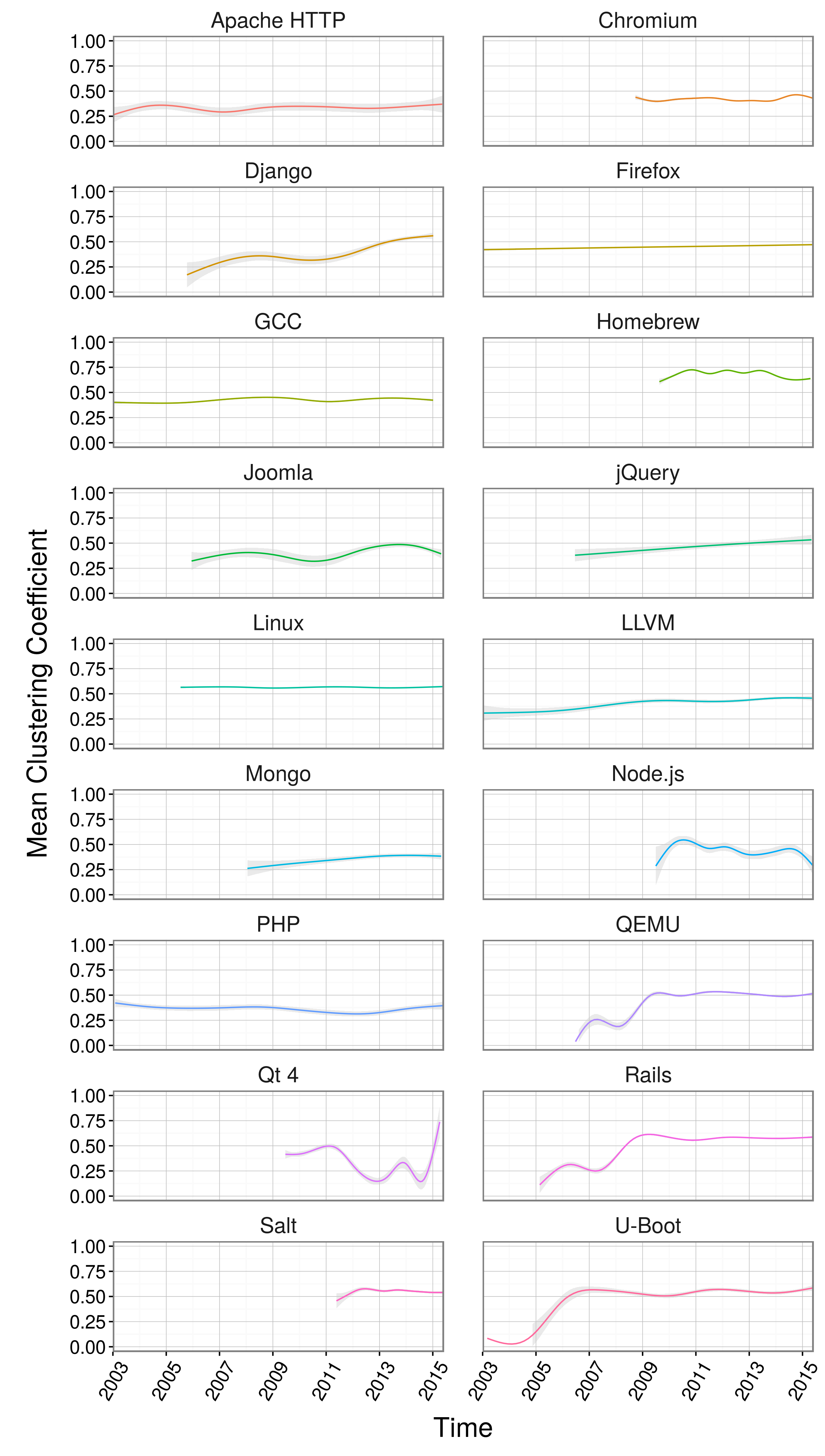}
\caption{Clustering-coefficient time series for all subject projects. The light gray area indicates the 99.5\% confidence interval.}
\label{fig:clustering_coefficient_ts}
\clearpage
\end{figure}

The clustering coefficient is
a means to measure the extent to which developers form cohesive groups.
In Figure~\ref{fig:clustering_coefficient_ts},
we present the evolution of developer networks with respect to their clustering coefficients for all subject projects.
The evolution of each project is illustrated by a time series that represents the mean clustering coefficient with a
light gray boundary to indicate the 99.5\% confidence interval. 
There is one evolutionary profile, in particular, that describes the majority of the subject projects. This profile is characterized by a positive trend component (i.e., increasing clustering coefficient)
that smoothly converges to a clustering coefficient range of 0.45--0.55. For the projects that do not fit this profile, the positive trend component is not observable,
possibly because we do not have a complete project history. For example, the development of the Linux kernel was started in
1996, but the publicly available Git repository only has commits dating back until early 2005. In Table~\ref{table:project_measurements}, we present the results of the mean clustering coefficients (column ``$\mu_{cc}$'') and variances (column ``$\sigma_{cc}^{2}$'') for the most recent revisions of each project.

\begin{figure}[t]
\centering
\includegraphics[width=0.8\linewidth]{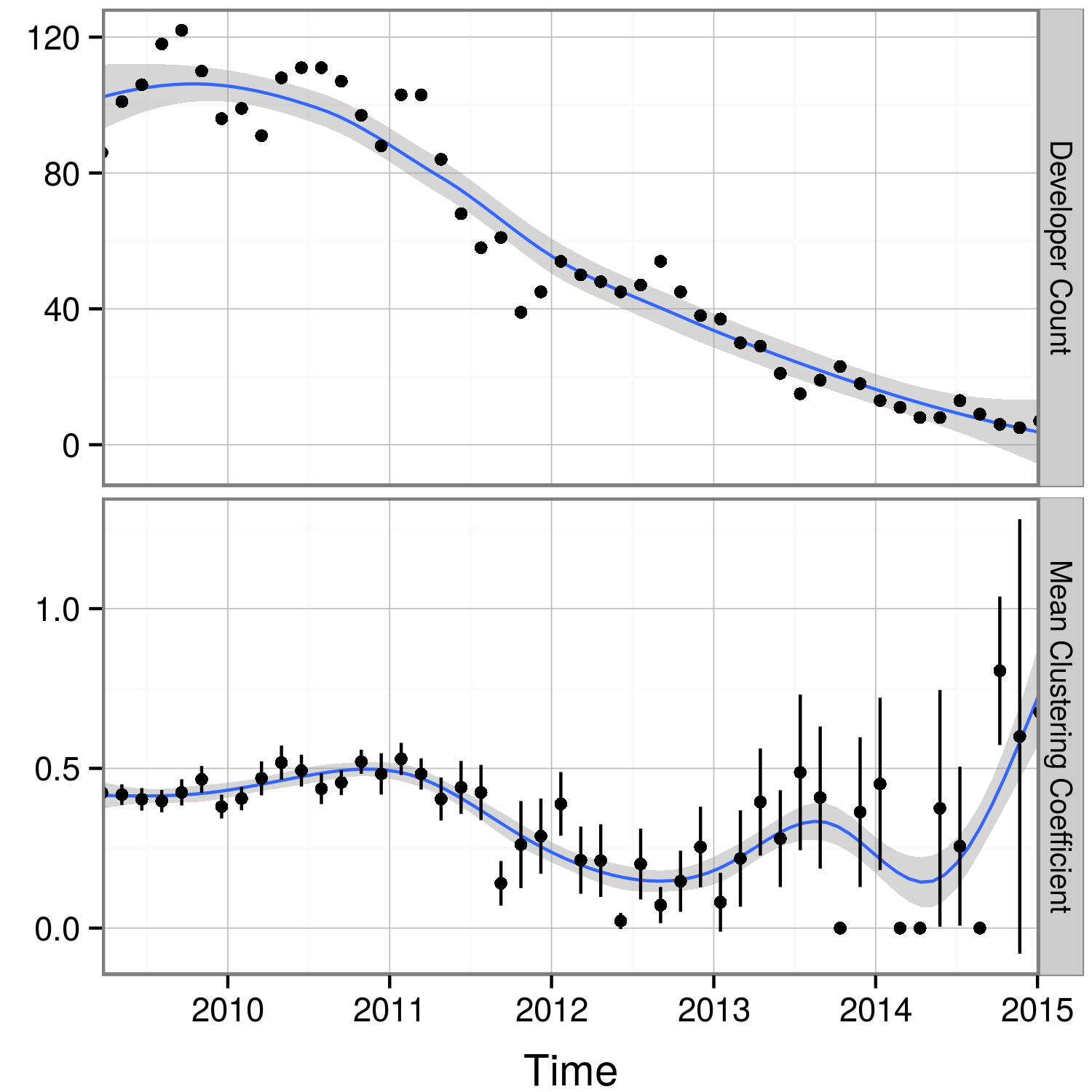}
\caption{Developer count and mean clustering coefficient evolution for Qt with the standard error bars for the mean clustering coefficient included. The significant decline in developer count co-occurs with instability in the clustering coefficient, indicating that departing developers have a significant impact on the local connectivity of the developer network.}
\label{fig:qt_growth}
\end{figure}

The only project which does not closely adhere to the general pattern of convergence is Qt. The exceptional behavior
seen in Qt is likely a consequence of the significant decrease in the number of active developers and possibly represents an evolutionary anti-pattern. The number of developers contributing to Qt is high until 2011 (see Figure~\ref{fig:qt_growth}), but this period is followed by several years of rapid decline. Similarly, we see that the mean clustering coefficient profile fits the general pattern until 2011, where the value suddenly drops and then oscillates before a radical upswing. It is worth noting that Qt is the only subject project
exhibiting this pattern, and it is similarly the only subject project that
has had such significant decline in number of contributing developers. 

For the majority of the projects, the fact that they do not ever significantly exceed a clustering coefficient of 0.55 suggests that there is a limitation to the distribution of coordination requirements in the local developer neighborhood. To give a reference point, a clustering coefficient of 0.5 for a node means that there are edges between half of the nodes neighbors. We observe that there is a preference to achieve a state where roughly half of every developer's neighbors also have a coordination requirement (i.e., an edge). The evolutionary profiles of our subject projects indicate that developer networks evolve according to a process that promotes coordination requirements increasing up to a maximum, but that prevents the formation of coordination requirements between too many developers.
There appears to be no non-zero lower bound on the clustering coefficient, but in all cases, an initially low clustering coefficient
does tend to converge towards a clustering coefficient of 0.5. The implication of a bound is that no observation should ever violate the bound by crossing it. In this sense, we are not able to prove that the upper limit that we observed will never be violated. The statement that we are able to make at this point is that the observations made on these 18 subject projects are evidence that a bound may exist. Particularly, the smooth convergence (i.e. gradual decrease in first derivative) to this upper limit is indicative of a bound that is not likely to be crossed in the future. 

\begin{figure}[t]
\centering
\includegraphics[width=0.8\linewidth]{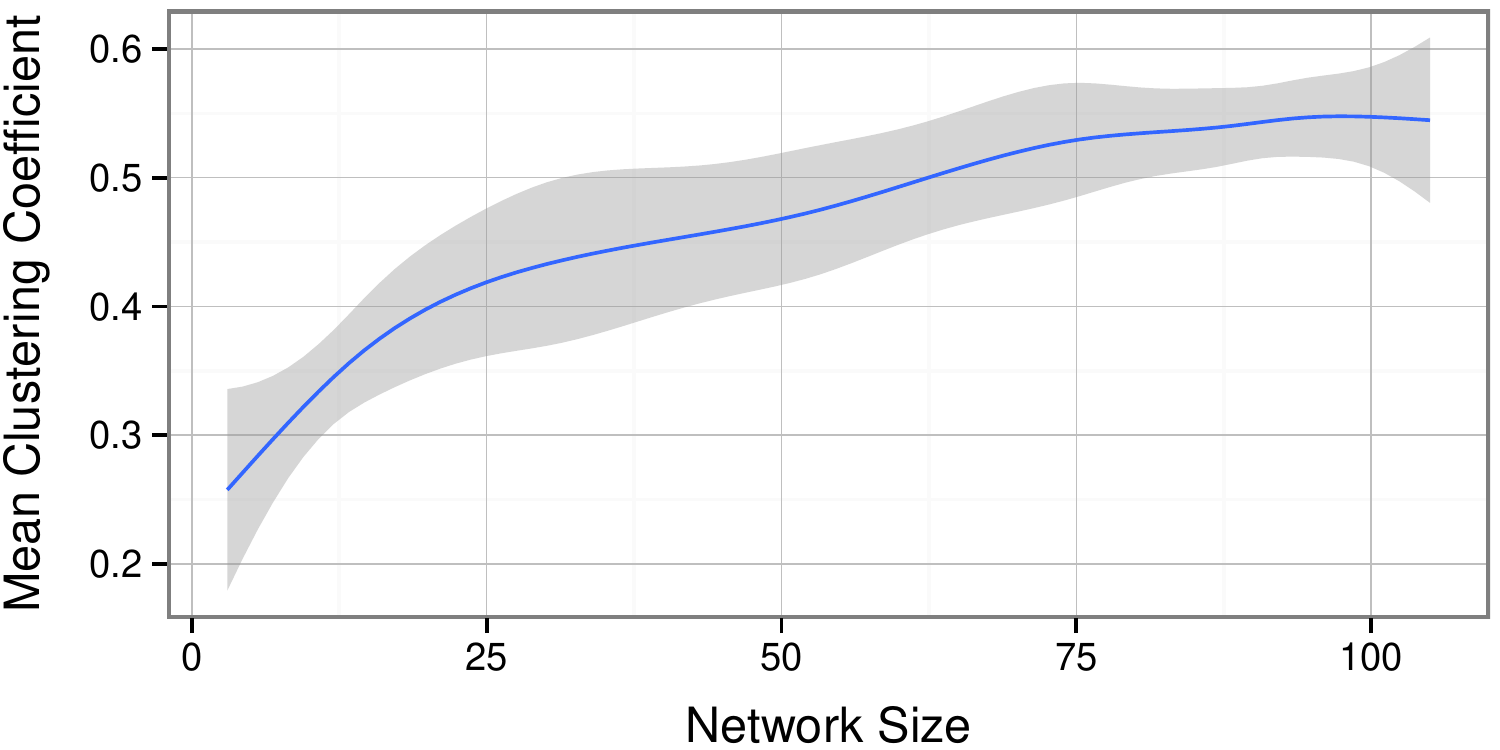}
\caption{Clustering coefficient versus network size for the history of Django, with a light gray boundary to indicate the 99.5\% confidence intervals.}
\label{fig:clustering_coefficient_vs_size}
\end{figure}

To better understand the mechanism behind the tendency for developers to form cohesive groups, we examined the relationship
between network size (i.e., number of developers) and clustering coefficient. In all cases, we found that the clustering coefficient
increases with the network size, however, this dependency decreases as the network size increases. This relationship
is shown for subject project Django in Figure~\ref{fig:clustering_coefficient_vs_size}, which illustrates a roughly logarithmic relationship
between network size and clustering coefficient. This is a notable result because, in the ER random graph described in Section~\ref{sec:scale_free_networks},
the clustering coefficient decreases with network size, and in many real-world networks, clustering coefficient and network size are independent~\citep{albert2002}.
From this observation, we conclude that developer networks form groups according to a non-random organizational principle that is also different from the preferential-attachment model used to explain many real-world scale-free networks. 

\vskip 1ex
\noindent\fbox{\parbox{\linewidth}{\ It has been hypothesized that, at a critical upper bound, the cost incurred from the overhead of coordination exceeds
the benefit of coordinating~\citep{Brooks1975}.
Our results indicate 
that this bound indeed does exist and that developer coordination is constrained to evolve
in a manner that promotes groups to form but not to exceed an upper bound. The evidence shown here is not definitive proof of a bound, but it is supportive of the conjecture that a bound exists.
Thus, we \emph{accept H3}.}}

\subsection{Hierarchy}
\begin{figure}[t]
\centering
\includegraphics[width=0.8\linewidth]{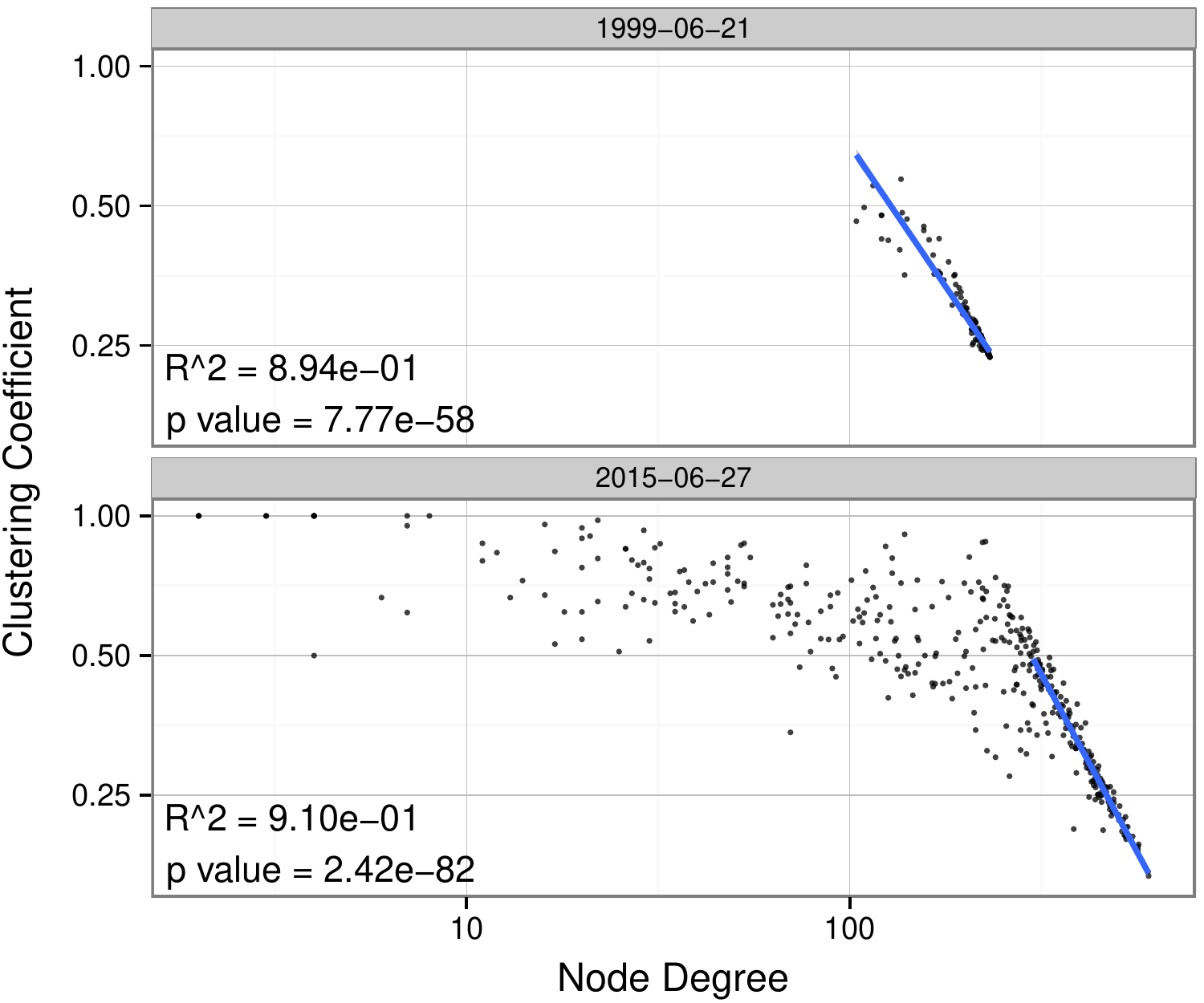}
\caption{Early and late stage hierarchy of Firefox. The fitted linear model is superimposed on a scatter plot of node degree versus clustering coefficient. In the early stage (top), the linear model describes the complete data set indicating global hierarchy. In the late stage (bottom), global hierarchy is not present since only the high-degree nodes are described by the linear model. The linear model in the late stage
has been fitted only to the high degree nodes.}
\label{fig:hierarchy}
\end{figure}

In Section~\ref{sec:network_hierarchy}, we introduced the concept of hierarchy
in terms of its relation to 
scale freeness and modularity: hierarchy is mathematically defined
by a linear dependence between the log-transformed clustering coefficient and the node degree.
We illustrate the results of applying the method described in Section~\ref{sec:network_hierarchy} in Figure~\ref{fig:hierarchy}, where the evolution of
hierarchy in an early stage (top) and late stage (bottom) is shown for Firefox. In the early state, we are able to see that the network
exhibits global hierarchy, because a linear model of the form $Y = \beta_{0} + \beta_{1}X$ describes the observed data, where $X$ is the degree of a developer and $Y$ is the clustering coefficient. In Section~\ref{sec:network_hierarchy}, we showed that a hierarchical network has the property that
the log-transformed degree and log-transformed clustering coefficient exhibit a linear dependence. The results indicate that the linear model achieves a good fit, which is evident by an $R^{2}$ value of 0.894. Furthermore, the $p$ value indicates that the linear model slope parameter $\beta_{1}$ is significantly different from zero, and so we can conclude that global hierarchy is present.
In the late stage (bottom figure), the linear model no longer describes the \emph{global} set of developers, instead it only describes the high degree nodes. In this case, we can conclude that a global hierarchy is no longer present and hierarchy predominantly exists in the high degree core developer group.

\begin{figure}[h!]
\centering
\includegraphics[width=0.8\linewidth]{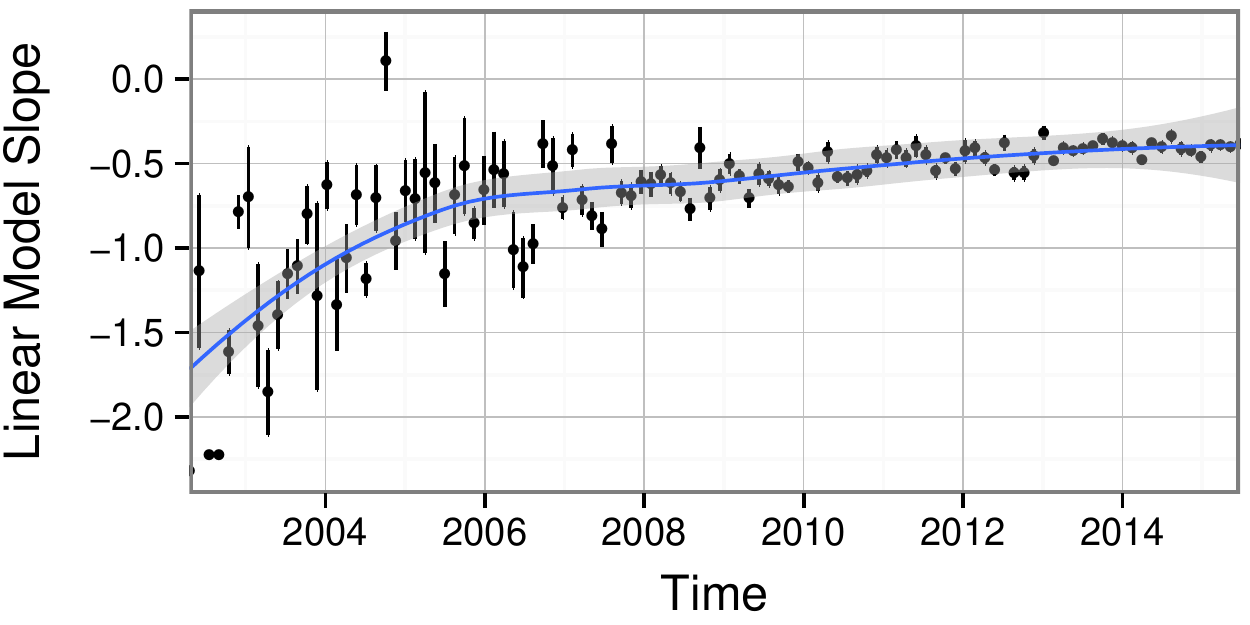}
\caption{Evolution of hierarchy for the entire history of LLVM. The gray area indicates the 99\% confidence interval, and error bars denote the standard error on the slope estimate. The trend indicates that hierarchy is decreasing over time as the linear model slope $\beta_{1}$ tends towards zero.}
\label{fig:hier_evolution}
\end{figure}

The principal evolutionary trend with respect to hierarchy is the following: In early
stages, developers are arranged in a global hierarchy. In later stages,
a hybrid structure emerges, where only the core developers are 
hierarchically arranged, but the global hierarchy is no longer present.
We observed that there is a smooth transition between the early and late stages,
shown for Firefox in Figure~\ref{fig:hierarchy}, which leads to a gradual
deconstruction of the global hierarchy.
The gradual deconstruction process is shown in Figure~\ref{fig:hier_evolution},
where we illustrate the continuous evolution of hierarchy over the entire history of LLVM.
Each sample represents the slope parameter $\beta_{1}$ of the
linear model describing the hierarchy in the project at a single point in time.
We see that hierarchy is most significant (largest negative slope) at the start and
is progressively lost until virtually no global hierarchy is present (i.e., near zero slope)
in the most recent revision. The results for all projects are shown in Table~\ref{table:project_measurements},
where column ``$\beta_{1_{early}}$''
represents the linear-model slope parameter at an early stage and column ``$\beta_{1_{late}}$''
represents the slope parameter at a late stage. 
The early stage represents the earliest analysis window with more than five developers present,
and the late stage represents the most up-to-date analysis window.
We are able to see that, in
all projects except one (Qt 4), $\beta_{1_{early}} < \beta_{1_{late}}$ indicating
that the hierarchy has diminished over time.

The results certainly suggest that, from a global perspective,
developer hierarchy diminishes with time,
but the mechanism responsible for the transformation is not obvious.
To investigate this process further, we examined the high-degree nodes (i.e., core developers)
and found that they are hierarchically arranged at all times.
Furthermore, the mechanism for decomposing the global hierarchy
is established through the introduction of low-degree and mid-degree nodes, which do not
obey the hierarchy established by the high degree nodes.
In essence, the developers become divided into two high-level organizational
structures: The highest-degree nodes (core developers) are hierarchically arranged and the mid-to-low-degree
nodes (peripheral developers) are not hierarchically arranged. This is visible in the late stage scatter plot
of Figure~\ref{fig:hierarchy}, where beyond the break point (at a degree of roughly 250), the nodes
obey a linear dependence and are thus hierarchically arranged. This evidence
suggests that the differences between core and peripheral developers are not entirely explained by
their distinct participation levels, but they are also distinct in how they are structurally embedded
in the organization.
\vskip 1ex
\noindent\fbox{\parbox{\linewidth}{\ As a project grows and becomes more complicated along numerous dimensions, we expected changes in the command-and-control structure. The expected trend was towards greater distribution of influence, which would manifest as the elimination of a global hierarchy. Our results indicate that, over time, hierarchy vanishes by the introduction of a large number of low and mid-degree developers that do not arrange into a hierarchy. We did also find evidence that extremely high-degree developers remain hierarchically arranged over time, though, they only constitute a very small faction of the project's developers. Overall, we \emph{accept H4}, because the hypothesis is a statement regarding the global structure and, in that sense, the evidence indicates that global hierarchy vanishes over time.}}

%\begin{figure}[htb]
%\includegraphics[width=\linewidth]{all_projects_hier_slope_time.png}
%\caption{Slope of clustering coefficient vs. degree for all subject projects. The light gray boundary indicates the 99\% confidence interval.\TODO{remove}}
%\label{fig:hier_all}
%\end{figure} 

\section{Threats to Validity}
\label{sec:threats_to_validity}
\paragraph{External Validity.}
We draw our conclusions from a manual selection of 18 open-source software projects.
The manual selection and the choice to analyze only open-source software projects is a threat
to external validity. We mitigated the consequences by choosing a wide variety
of projects that differ in many dimensions and constitute a diverse population.
Furthermore, we considered the entire history to prevent temporally biasing our results. 
We specifically chose only large projects with very active histories because our contributions
are focused on understanding complexity in developer coordination, and in small projects
(e.g., less than 10 developers), the coordination challenges are less severe. Due to certain limitations of our current analysis infrastructure, we are unable to include complete analyses of software ecosystem projects, such as Eclipse\footnote{\url{https://eclipse.org/}}, because they are typically distributed among multiple repositories.

\paragraph{Internal Validity.}
We examined the evolution of developer networks over time, however, it is conceivable that
factors other than time have an influence on the observed trends, threatening internal validity.
By considering the influence of network size,
we accounted for the most likely confounding factor.
Furthermore, we found that the trends are often consistent
across several projects, and we rigorously employed statistical methods
to avoid drawing conclusions from insignificant fluctuations in the data. 

\paragraph{Construct Validity.}
Our methodology relies, to some extent, on the integrity of the data in the version-control system to generate a valid
developer network, threatening construct validity.
Since the version-control system is a critical element of the software-engineering process,
it is unlikely that the data would be significantly corrupt. In terms of network construction,
the heuristics we rely on have been shown to generate authentic developer networks,
but do omit some edges~\citep{joblin2015}. However, a few omitted edges would not severely
impact the conclusions of our study. Furthermore, our enhancement of this form of developer networks to recover omitted
edges is based on a technique that has been shown to authentically represent system coupling~\citep{bavota2013}. The operationalizations of scale freeness, modularity, and hierarchy
are thoroughly studied and well-established concepts in the area of network analysis. We further relied
on the degree of nodes to operationalize the concept core and peripheral developers.
Although the application concepts from social network analysis to socio-technical developer
networks is relatively new, empirical evidence is accumulating that suggests the metrics are reliable and valid~\citep{Meneely2011,joblin2015,joblin2016}.

\section{Discussion \& Perspectives}
The results of our study on the evolution of developer networks revealed several intriguing
patterns. We will now discuss the relevance and potential explanation for these network patterns
by linking them to software-engineering principles. Specifically, we discuss
a likely model for growth of a project, a source of pressure for developers to become
more coordinated with time, and the benefits of a hybrid organizational structure that is
hierarchical for core developers and non-hierarchical for peripheral developers.

\paragraph{RQ1:} Our first research question asked what evolutionary adaptations are observable in the evolution of developer networks. In terms of scale freeness, we saw that projects do not begin in a scale-free state, instead this property emerges over time. Initially, the structure of developers exhibits high homogeneity and then, over time, hub nodes (i.e., very involved developers) appear that are responsible for a disproportionately large number of coordination requirements. We also saw that adaptations occur in the modularity of developer networks. Developers are loosely clustered initially, but, over time, clustering increases and gradually converges to a state where half of every developer's neighbors have coordination requirements. Finally, the structural property of hierarchy also changes over the course of time. Initially, projects have a globally hierarchical organization. Over time hierarchy is lost, as peripheral developers are introduced to the network, which do not assimilate into the hierarchy. Still, hierarchy is always maintained for the highly-connected core developers. 

\paragraph{RQ2:} Our second research question focused on the relationship between these changes in the organizational structure and the scale of a project. Most of our subject projects experience steady growth over time. Presumably, many of the evolutionary principles we observed are closely related to the increasing scale of the project. For a couple of projects, we saw the growth stagnate or the overall size decrease. In these projects, we saw the reverse of what was seen during project growth. For example, the scale-free property was lost and clustering decreases. At this point, our results suggest a strong dependence between the scale of a project and the properties of its organizational structure. It appears that projects of different size exhibit different structural features of the organization. This result is interesting to the general software-engineering community, because it suggests that, when determining how to organize developers,
it is crucial to consider how many developers will be involved in the implementation. In projects
with few developers (e.g., less than 30), it may not be necessary to have developers that are highly dedicated to coordinating the work of others, and each developer can essentially occupy equivalent structural positions in the organization. However, in a very large project (e.g., 30 or more), it may be crucial to have developers entirely dedicated to coordinating the work of other developers and to occupy hub positions that span the organizational structure.

%Scale freeness
\paragraph{Scale freeness:} To better understand the growth behavior
of developer networks, we examined the relationship between a project's growth state (increasing or decreasing)
and the scale-freeness property of its developer network. The model of preferential attachment, which is the
predominant generative model for scale-free networks, has the simultaneous 
requirements that the network must grow and that new nodes have a preference
to attach to already well-connected nodes~\citep{barabasi1999}.
We found that, in this regard, the evolution of developer networks into a scale-free state is consistent with the model for preferential attachment. For several projects, the scale-freeness property is only observable
during network growth and is lost during periods of growth stagnation or
decrease, as shown for Node.js in Figure~\ref{fig:growth_profile}. Furthermore,
the loss of the scale-freeness property often precedes the stagnation or loss of developers. While
it would be premature to make any strong statement about causality, the combination of correlation
and preceding in time makes the loss of the scale-free state a conceivable predictor for the
loss of growth in the project. These results suggest that, if a project grows beyond
a certain size, the coordination structure will exhibit strong inhomogenity in the distribution of coordination requirements among developers. It seems that there is
a driving force that encourages a relatively small group of developers to bear the
majority of the coordination burden. As a project achieves a large size (50 developers or more), this need for hub nodes in the coordination structure appears to be more critical. Software engineers should consider
the project size when determining how to distribute the tasks among developers and how modes of collaboration between the multiple development sites should be realized.

%Modularity
\paragraph{Modularity:} In Section~\ref{sec:results_modularity}, we noted that an increasing clustering coefficient
is a common evolutionary trend. This is a curious result because, in the ER random graph model
(see Section~\ref{sec:scale_free_networks}), the clustering
coefficient decreases with increasing network size, while in the preferential-attachment model, the clustering
coefficient is independent of network size~\citep{ravasz2003hierarchical}. 
So, this result begs the question of what the driving force behind this unique evolutionary trend is.
From the theory of software evolution, we
expect that the natural tendency for an architecture is to become more strongly coupled over time, as
complexity increases and initially clean abstraction layers deteriorate~\citep{lehman2001}, which
has been observed also in practice~\citep{mens2008evolution}. Additionally, Conway's law
suggests that the organizational structure and the structure of technical artifacts produced by the organization
are constrained to mirror each other~\citep{Conway1967}. Based on these principles, we hypothesize that the
evolution of the artifact structure is the driving force that influences developers to become more coordinated.
Software engineers should be conscientious of the increasing demand on developers to
coordinate with more developers as the software evolves. In the later stages of a project, it may be critical to shift
more attention towards mechanisms that support effective coordination
between developers. It may even be necessary to reduce the task load on developers
in later stages of a project, to ensure that the coordination requirements receive
sufficient attention; otherwise a decrease in software quality is a legitimate threat.
%We think that this dynamic is not present in other real world networks because the product of the coordination
%does not influence organization as strongly as in software development. For example, in the scientific authorship
%network, it is conceivable that the motivations to collaborate are predominantly external to the
%the structure of scientific knowledge (i.e., the product of collaboration) and are more directly
%influenced by co-location or shared funding opportunities.
%In contrast, a software project may require two developers to coordinate simply because the architecture demands it,
%regardless of all other factors.

%Hierarchy
\paragraph{Hierarchy \& Stability:} One of the most intriguing characteristics of open-source software projects is the strongly inhomogeneous distribution
of effort between core and peripheral contributors~\citep{koch2004,Toral2010,Fielding2000}.
This characteristic is distinct from typical commercial development setups and is conceivably responsible for enabling 
open-source software projects to scale without reducing overall productivity, which violates conventional software-engineering wisdom~\citep{koch2004}.
Typically, core and peripheral developers are classified based on the number of commits, lines of code,
or e-mails they contributed~\citep{joblin2016}. Interestingly, we discovered that the differences between the two groups
are also observable in their organizational structure and stability, where the group of core developers is both hierarchically organized and relatively stable, but
the group of peripheral developers is both unstable and not hierarchically organized. We think that the reason for peripheral developers not assimilating into the
hierarchy stems from pressures to form a hybrid organizational structure that promotes regularity while also remaining
flexible. The process of software development demands a high degree of consistency and, for this reason,
hierarchies are appropriate organizational structures. However, hierarchies are intrinsically inflexible
structures~\citep{kotter2014}. In open-source software projects, there is pressure for the organizational structure to remain flexible
because, as we have shown, open-source software projects have high developer turnover rates for the peripheral developers, who constitute the majority of the contributors. It is conceivable that the existence of a hybrid organizational
structure is even a signal of project health by indicating that the organization has responded
to the adaptation pressures that are present in open-source software development. To the wider software-engineering community,
this result indicates that a software project may benefit from embedding developers into the organizational
structure differently depending, on their experience level and likelihood of leaving the project.
For example, a hierarchy can be an efficient structure when the members of the hierarchy a not likely
to exit the hierarchy. In the same way that open-source software development avoids embedding the volatile peripheral developers
into the hierarchy composed of core developers, it may offer benefit for any software project to avoid integrating inexperienced or potentially volatile developers into their hierarchical organizational structure.

\section{Related Work}
%network construction
Lopez et al.\ first studied developer coordination by linking developers based on mutual contributions to
modules for a static snapshot of three open-source software projects. They found that developer networks are not scale free, based on a
visual inspection of the cumulative degree distribution~\citep{Lopez-fernandez2006}.
Jermakovics et al.\ constructed networks based on contributions to files for three projects, and they developed
a graph-visualization technique to represent the developer organizational structure~\citep{Jermakovics2011}. Toral et al.\ constructed
developer communication networks based on the Linux kernel e-mail archives between 2001 and 2006~\citep{Toral2010}. They found that
participation inequality is present in the communication network, and they introduced a core--peripheral-developer
classification scheme. We differentiate our work by analyzing the entire project history and viewing
developer coordination as an evolutionary process. Our network-construction procedure has demonstrated valid results with respect
to capturing developers' perception of who they collaborate with and reveals a statistically significant community structure, which
is obscured by the more coarse-grained approaches used in prior work~\citep{joblin2015}.
Additionally, we use a fully automated and statistically rigorous framework to reduce subjectivity, and we draw our conclusions from 18 projects
instead of just two or three. We build on prior work by explaining the commonly observed network features (e.g., participation inequality) in terms of the important structural concepts of scale freeness, modularity, and hierarchy. 

%network structure
Louridas et al.\ studied structural dependencies
between classes and packages of 9 software systems using static source-code
analysis techniques~\citep{louridas2008}. They found that
power-law distributions are a ubiquitous phenomenon in the dependency structure by fitting
a line to the log-scale degree distribution. Our work is complementary by
identifying power-law distributions in developers' coordination requirements. This is a step towards
an empirical validation of Conway's law by showing that a necessary condition is met
regarding the match between the organizational structure and technical artifact structure.

%Developer turnover
While there is a number of theories regarding developer turnover and its effects, current empirical results are limited. Foucault et al.\ examined the relationship between internal and external developer turnover on software quality in terms of bug density~\citep{foucault2015}. Consistent with current theories, they found that high external turnover has a negative influence on module-level bug density.
Others have explored factors that contribute to developer turnover and motivations for long-term involvement~\citep{Yu2012,Hynninen2010,Schilling2012}. Mockus found that developers leaving projects negatively influence code quality, while new developers entering the project have no influence~\citep{Mockus2010}. Oddly, the results of Mockus and Foucault et al.\ do not agree, which may suggest that the influence of turnover is dependent on additional context factors. In our work, we primarily focused on the relationship between the turnover characteristics of core and peripheral developer groups and how these distinct groups are structurally embedded in the organization. We use the distinct turnover rates to rationalize the evolution of the developer network as an optimization process.

%network evolution
%Large-scale investigations of software evolution are scarce,
%and longitudinal studies are expensive, because the historical data may be unavailable
%and the time varying nature presents added challenges. 
Godfrey et al.\ were the first
to study software evolution in open-source software and found that the Linux kernel violates
principles of software evolution by achieving super-linear growth at the system level~\citep{godfrey2000}.
This was later supported by evidence extracted from the version control system of
8621 projects on SourceForge.net~\citep{koch2004}. Koch found that large open-source software projects violate 
several laws of software evolution established for commercial projects~\citep{koch2004}.
Specifically, they showed that developer productivity is independent
of the number of developers in the project---a direct violation of Brooks' law~\citep{Brooks1975}.
Furthermore, participation inequality is common and increases with the system size---a result that we
confirmed---but the increase in inequality does not influence developer productivity.
Koch proposed that strict modularization, self-organization, and highly decentralized work are responsible
for the high efficiency seen in open-source software projects, but this was never verified~\citep{koch2004}. 
In our study, we found that
our more detailed methodology, which considers source-code structure and software coupling, supports
prior observations. Furthermore, we were able to extend the body of knowledge by directly studying
the evolution of coordination structures that are conjectured to be responsible for the remarkable
properties of open-source software projects.

%similar observations in other fields
%In a different, but nonetheless related field, researchers have shown that a network
%representing the collaborative organization of knowledge in Wikipedia,
%based on references between articles, is a scale-free network and the network
%growth is described by the law of preferential attachment~\citep{spinellis2008}.
%By showing that power laws are present in OSS project, we add to the evidence
%that healthy collaborative enterprises are ubiquitously scale-free.

% TODO: Other possible related work
%- Striking a Balance Between Trust and Control in a Virtual Organization: A Content Analysis of Open Source Software Case Studies
%- A Case Study of Open Source Software Development: The Apache Server 
%- Two Case Studies of Open Source Software Development: Apache and Mozilla 
%- Effort, Cooperation and Coordination in an Open Source Software Project: Gnome}

\section{Conclusion}
From an organizational perspective, open-source software projects are an extreme example of
large-scale globally-distributed software engineering, and as such, represent a unique opportunity
to study developer-coordination mechanisms. Despite the lack of mandated organizational
structures, we found that open-source software projects are constrained to evolve according to non-random
organizational principles.

Extracting and processing the operational data stored in the version-control system proved to be challenging.
By pairing our network-construction procedure with a sliding-window technique, we were able to
identify important insights that would be hidden in a temporally static view.
In addition, we enhanced the developer networks by using information retrieval approaches to gain a more rich view of the coordination requirements that exist between developers.
We see these technical contributions as meaningful steps towards advancing the techniques for mining software repositories.

Based on our longitudinal study of 18 open-source software projects, we found that, in projects exceeding 50 developers,
the coordination structure becomes scale free.
We also found that there is a tendency for an increasing number of coordination requirements to appear among groups of developers, but the increasing trend is likely limited by a particular upper bound,
where coordination requirements exist between roughly half of every developers neighbors.
Additionally, we discovered that developers are hierarchically arranged in the early phases of a project, but in later phases the global hierarchy vanishes and a hybrid structure emerges, where core developers
form the hierarchy and peripheral developers exist outside the hierarchy. With this result, we demonstrated that core and peripheral developers---which are traditionally defined based on their level of participation---also differ in how they are structurally embedded in the project's coordination structure. Overall, the adaptations that we observed in the structural features balance the opposing constraints of supporting effective coordination and achieving robustness to developer withdrawal. Finally, we discussed how these structural features enable a project in benefiting from a large, but volatile, peripheral developer group, while at the same time, supporting effective coordination and regularity between the much more stable core developer group. From these results, it is clear that significant structural changes occur in
the coordination structure of a project over time, and particularly as developers are added.
These insights provide valuable information to software engineering practitioners by highlighting the
impact that adding developers has on the coordination structure. With this knowledge we can begin
to establish strategies for integrating new developers that try to minimize the disruption
to the existing coordination structure.

Apart from the general patterns that explain the majority of subject projects, we also noted a number
of interesting deviations from the general patterns. For example, during a period of time, when Firefox was experiencing notable project delays and turmoil within the developer community, we observed that
scale freeness suddenly disappeared. In Node.Js, there was an oscillatory behavior to the number
of contributing developers and the scale-freeness property was lost whenever the project was not in a
growing state. Finally, for the few projects that never became scale free, or only for a brief time,
a developer group larger than 60 was never sustainable and the number of contributing developers decreased
shortly after reaching a maximum, as was the case for PHP, jQuery, and Apache HTTP.
It was often the case that projects deviating significantly from the general patterns were experiencing
other negative project conditions such as significant loss in the number of active developers.

Since large-scale studies of software evolution are rare, and studies on the coordination
structure are even more rare yet, we see our work as an important step towards understanding
the coordination mechanisms that are present in large-scale globally-distributed software engineering.
We hope that, by making our analysis infrastructure publicly available, 
we lower the barrier to contributing to this field of research and
accelerate the pace of progress.

\appendix\normalsize
\appendixpage
\section{\\Function-level Semantic Coupling}
To determine function-level semantic coupling,
we first extracted the implementation for each function in the system,
including all source code and comments.
We then employed well-established text-mining preprocessing operations with minor modifications for our specific domain requirements.
In this framework, each function is treated as a ``document'' in the text-mining sense of the word, and then the document collection was processed use the following processing operations.

\paragraph{Preprocessing.}
The preprocessing stage primarily focuses on reducing word diversity and elimination of words that contain little information. \emph{Stemming} is to used to reduce words to their root form by removing suffixes (e.g., ``ing'', ``ly'', ``er'', etc.) from each word in the document. Stemming is necessary because, even though a root word may have several forms by adding suffixes, it typically refers to a relatively similar concept in all forms. In software engineering, there is a number of variable-naming conventions, such as letter-case separated (e.g., CamelCase) or delimiter separated words that need to be tokenized appropriately. We added additional preprocessing stages to specifically handle proper tokenization of popular naming conventions. For example, the function identifier ``get\_user'' or ``getUser'' are separated into the two words ``get'' and ``user''. One simple example of why this is important is that getters and setters interacting with the same attribute would be incorrectly understood as distinct concepts without appreciating the variable-naming conventions. The final stage of the preprocessing is to remove words that are known not to contain useful information based on a-priori knowledge of the language. For example, words such as ``the'' are not helpful in determining the domain concept of a document. Removing these words is beneficial for the computational complexity and results by reducing the problem's dimentionality and attenuating noise in the data.

\paragraph{Term Weighting.}
After the preprocessing stage, we arrange all remaining data into a term--document matrix, for mathematical convenience. A term--document matrix is an $M\times N$ matrix with rows representing terms and columns representing documents. For example, an element of the term--document matrix $\mathit{TD}_{i,j}$ is non-zero when document $d_{j}$ contains term $t_{i}$. All elements of the term--document matrix are integer weights that indicate the frequency of occurrence of a given term in a given document. We then apply a weight transformation to the term--document matrix based on the statistics of occurrence for each term. Intuition suggests that not all terms in a document are equally important with regard to identifying the domain concept. The goal of the weighting transformation is to increase the influence of terms that help to identify distinct concepts and that decrease the influence of the remaining terms. The particular weighting scheme we applied is called \emph{term frequency-inverse document frequency}:
\begin{equation}\label{eq:weighting}
\textit{tf-idf}_{t,d} = \textit{tf}_{t} \times \log\frac{N}{\textit{df}_{t}}.
\end{equation}

\noindent
The term $\mathit{tf}_{t}$ represents the global term frequency across all documents. The second term is the logarithm of the inverse document frequency, where $N$ is the number of documents in the total collection and $\mathit{df_{t}}$ is the number of documents that term $t$ appears. Upon closer inspection, one can recognize that Equation~\ref{eq:weighting} is: (a) greatest when a term is very frequent, but only appears in a small number of documents, (b) lowest when a term is present in all documents, and (c) between these two extreme cases when a term is infrequent in one document or occurs in many documents. 

\paragraph{Latent Semantic Indexing.}
Even for a modest-sized software project, the number of terms used in the implementation vocabulary easily exceeds the thousands. The problem with this becomes evident when adopting the vector-space model, where we consider a document as a vector that exists in a space spanned by the terms that comprise the document collection. Fortunately, this very high dimensional space is extremely sparse, which allows us to project the documents into a lower dimensional subspace, which makes the semantic similarity computation tractable. We achieve this using a matrix decomposition technique that relies on the singular value decomposition called \emph{latent semantic indexing}. An added benefit of this technique is that it is capable of correctly resolving the relationships of synonymy and polysemy in natural language~\citep{baeza1999modern}. Furthermore, latent semantic indexing has shown evidence to be valid and reliable in the software-engineering domain~\citep{bavota2013}. 

\paragraph{Semantic Similarity.}
In the final step of the analysis, we determine semantic coupling by computing the
similarity between all document vectors projected onto the lower dimensional subspace attained from applying latent semantic indexing. We operationalize the similarity between two document vectors in the latent space using cosine similarity

\begin{equation}
\text{similarity}(\vec{d_{a}}, \vec{d_{b}}) = \frac{\vec{d_{a}} \cdot \vec{d_{b}}}{\left \| \vec{d_{a}} \right \|\left \| \vec{d_{b}} \right \|},
\end{equation}
\noindent
where the numerator is the dot product between the two document vectors and the denominator
is the multiplication of the magnitude of the two document vectors. Intuitively, cosine similarity expresses the difference in the angle between the two document vectors; it equals 1, when the two vectors are parallel, and 0, if they are orthogonal. Two source-code artifacts
are then considered to be semantically coupled if the cosine similarity exceeds
a given threshold.
We experimented extensively with a number of thresholds by manually inspecting the results and judging whether the functions were, in fact, semantically related using architectural knowledge of a well-known project\footnote{\url{http://siemens.github.io/codeface/}}. We found that a threshold of 0.65 was able to identify most semantic relationships with only a very small number of false positives. We did, however, cautiously chose the threshold to optimize to avoid false positives rather than false negatives.

\section{\\Analysis Window Selection}
We chose to use a sliding-window approach in our study to generate the time-resolved series of developer networks. Another option would have been to analyze the project using non-overlapping windows, but this can lead to problematic edge discontinuities between the analysis windows. For example, a set of several related changes to the software could be divided between two different analysis windows, even though the changes occurred temporally close together. For this reason, a sliding-window approach superior to the alternative for our purposes, but we also recognized that overlapping windows could influence the appearance of developer transitions (see Section~\ref{sec:markov_chain}), because a commit can appear in two contiguous analysis windows. To test whether the overlapping windows distorts the overall outcome, we compared all Markov chains using non-overlapping windows with those using overlapping windows. The comparison revealed that, in all projects, our conclusion that core developers are more stable than peripheral developers is true regardless of which windowing strategy is used. In most cases, using non-overlapping windows increased the probability that a core or peripheral developer leaves the project, but peripheral developers are always significantly more likely to leave. For example, in QEMU, using overlapping windows, core and peripheral developers leave a project with 0.5\% and 10.9\% chance respectively. In the case of non-overlapping window this changes to 13\% chance for core and 55\% chance for peripheral. The data for both sets of Markov chains are included on the supplementary Web site.


\begin{acknowledgements}
This work has been supported by the German Research Foundation (AP~206/4, AP~206/5, AP~206/6).
\end{acknowledgements}

\bibliographystyle{spbasic}      % basic style, author-year citations
\bibliography{evolution_paper}   % name your BibTeX data base

\end{document}